\documentclass[12pt]{article}

\usepackage{amsmath}
\usepackage{amsfonts}
\usepackage{amssymb}
\usepackage{amscd}

\newcounter{theorem}
\renewcommand{\thetheorem}{\arabic{section}.\arabic{theorem}}

\newenvironment{thm}[1]{\par
\begin{sloppypar}\refstepcounter{theorem}%
\noindent{\bf #1 \thetheorem.}\it{}}{\end{sloppypar}}

\newenvironment{proposition}{\begin{thm}{Proposition}}{\end{thm}}
\newenvironment{corollary}{\begin{thm}{Corollary}}{\end{thm}}
\newenvironment{lemma}{\begin{thm}{Lemma}}{\end{thm}}

\newenvironment{defi}[1]{\par
\begin{sloppypar}\refstepcounter{theorem}%
\noindent{\bf #1 \thetheorem.}\rm{}}{\end{sloppypar}}
\newenvironment{definition}{\begin{defi}{Definition}}{\end{defi}}

\newcommand{\eh}{\hfill}\newlength{\sperr}
\newenvironment{proof}{{\settowidth{\sperr}{\rm Proof}
\par\addvspace{0.3cm}\noindent\parbox[t]{1.3\sperr}{\rm P\eh r\eh o\eh o\eh 
f\eh.}}}{\nopagebreak\mbox{}\hfill $\blacksquare $\par\addvspace{0.25cm}}

\def\R{{\rm I\kern-.2em R}}   
  \def\H{{\cal H}} 
   \def\Op{\mathfrak{Op}}

\begin{document}
\vspace*{3cm}
\begin{center}

\begin{LARGE}\textbf{The Magnetic Weyl Calculus}\end{LARGE}
\vspace{2cm}

\begin{Large}Marius M\u antoiu and Radu Purice\end{Large} \\
Electronic 
mail: Marius.Mantoiu@imar.ro, Radu.Purice@imar.ro
\vspace{2cm}

\textit{Institute of Mathematics ``Simion Stoilow'' of
the Romanian Academy}, \\
P.O.  Box 1-764, Bucarest, RO-014700, Romania, 

\end{center}
\newpage

\begin{abstract} 

In the presence of a variable magnetic field, the Weyl pseudodifferential calculus must be modified. The usual modification, based on ``the minimal 
coupling principle'' at the level of the classical symbols, does not lead to gauge invariant formulae if the magnetic field is not constant. We 
present a gauge covariant quantization, relying on the magnetic canonical commutation relations. The underlying symbolic calculus is a 
deformation, defined in terms of the magnetic flux through triangles, of the classical Moyal product.
\end{abstract}



\section*{Introduction} 

The correspondence principle of Quantum Mechanics asks that for a given physical system there should be a systematic way to convert 
classical observables into quantum observables. We use for this the rather vague term of {\it quantization}. For many given systems a true, 
manageable quantization is problematic, but there are important situations in which commonly accepted solutions exist. The purpose of the 
present paper is to propose what we think to be the correct solution for the case of a non-relativistic spinless particle, moving in $\mathbb 
R^N$, in the presence of a variable magnetic field. From the mathematical point of view, our results may be considered as a first step towards a 
quantization of symplectic manifolds.

General principles assert essentially that classical observables are functions in phase space, while quantum observables should be self-adjoint 
operators in some Hilbert space. Rather often, for some basic observables (positions, momenta,...) the prescription is either essentially unique 
(may be due to some commutation relations), or at least generally accepted. Thus, for many physical systems, quantization of all phase-space 
functions could be regarded as a sort of functional calculus. But since, as a rule, the basic observables do not commute, one cannot rely on the 
usual spectral theory to define this functional calculus. Roughly, quantization may be seen as the mathematical problem of defining functions of 
several non-commuting self-adjoint operators. Of course, the features of the physical system both impose constrains and offer empirical 
suggestions with respect to this procedure.

In the absence of any magnetic field, a non-relativistic spinless particle moving in $\mathbb R^N$ is quantized through the Weyl 
pseudodifferential calculus. If $f$ is a suitable function (``symbol'') defined on the phase space $\mathbb R^{2N}$, the corresponding operator is 
defined to act in the Hilbert space $L^2(\mathbb R^N)$ by the formula

\begin{equation}\label{prima}
[\mathfrak{Op}(f)u](x):=\int_{\mathbb R^{2N}}dy\;dp\;e^{i(x-y)\cdot p}f\left(\frac{x+y}{2},p\right)u(y).
\end{equation}

In a certain sense (which can be made precise and which will be discussed below), we may write $\mathfrak{Op}(f)=f(Q,P)$ and interpret it as 
the action on the symbol $f$ of the functional calculus associated to the family of operators $(Q_1,\dots,Q_N,P_1,\dots,P_N)$, where $Q_j$ is 
the multiplication by the $j$'th coordinate and $P_j:=-i\partial_j$. The well-known rules of commutation between these position and momentum 
quantum observables play a decissive role in determining the explicit formula above. They are thus also basic in deducing the explicit product rule 
$(f,g)\mapsto f\circ g$ and involution $f\mapsto f^\circ$ leading to $\mathfrak{Op}(f)\mathfrak{Op}(g)=\mathfrak{Op}(f\circ g)$ and 
$\mathfrak{Op}(f)^*=\mathfrak{Op}(f^\circ)$.

When a magnetic field $B$ is turned on, we are faced with the problem of modifying the formula for $\Op(f)$ in a way taking into account the 
presence of the magnetic field in a correct, physical way.  A mistaken procedure which appears from time to time in the literature is the 
following: One chooses a vector potential $A$ corresponding to the magnetic field ($B=dA$) and, by an (unjustified) application of the minimal 
coupling principle, one sets $\Op_A(f):=\Op(f_A)$, with $f_A(x,p):=f(x,p-A(x))$. This is ment to be the action on $f$ of the functional calculus 
associated with the family $Q_1,\dots,Q_N,\Pi_1,\dots,\Pi_N$, where $\Pi_j:=P_j-A_j(Q)$ is the $j$'th component of the vector potential. But the 
resulting formula 

\begin{align}\label{ei}
[\Op_A(f)u](x):=
&\int\limits_{\mathbb R^{2N}}dy\;dk\;e^{i(x-y)\cdot k}f\left(\frac{x+y}{2},k-A\left(\frac{x+y}{2}\right)\right)u(y)=\\
&=\int\limits_{\mathbb R^{2N}}dy\;dp\;e^{i(x-y)\cdot p}e^{i(x-y)\cdot A\left(\frac{x+y}{2}\right)}f\left(\frac{x+y}{2},p\right)u(y)
\end{align}
cannot be the right one, since it lacks gauge covariance: If one chooses another vector potential $A'$ associated to $B$, differing from the initial 
one by the gradient of a scalar function, $A'=A+\nabla\rho$, then {\it the expected formula} $e^{i\rho}\Op_A(f)e^{-i\rho}=\Op_{A'}(f)$ {\it does not 
hold}. In fact, this formula is restored if one replaces the phase factor $e^{i(x-y)\cdot A\left(\frac{x+y}{2}\right)}$ by $e^{i(x-y)\cdot 
\int_0^1dsA((1-s)x+sy)}$. It is reassuring to note that $-(x-y)\cdot\int_0^1dsA((1-s)x+sy)$ is in fact the circulation $\Gamma^A([x,y])$ of the 
vector potential $A$ through the segment leading from $x$ to $y$. Thus the formula we propose instead of (\ref{ei}) is 

\begin{equation}\label{noi}
[\mathfrak{Op}^A(f)u](x):=\int_{\mathbb R^{2N}}dy\;dp\;e^{i(x-y)\cdot p}e^{-i\Gamma^A([x,y])}f\left(\frac{x+y}{2},p\right)u(y).
\end{equation}

The main purpose of our article is to give an explanation of these facts. As a prologue for doing this, in the first Section we review some facts 
related to canonical commutation relations and Weyl calculus when no magnetic field is present. The main topic will be the justification of the 
formula for $\Op(f)$ as a sort of integrated form of {\it the Weyl system}, which is a family of unitary operators $\{W(\xi)\}_\xi$ indexed by the 
points of the phase space and containing the relevant information on the commutation relations between the operators $Q$ and $P$. We claim no 
originality (see for example \cite{Hormander1}); we include this here because it seems to be an argument largely ignored, which is basic to our 
approach. The symbolic calculus beyond the Weyl prescription is the famous Moyal product. Other references emphasizing the connection 
between Quantum Mechanics and pseudodifferential theory are \cite{Grossman} and \cite{Folland}.

When a magnetic field is present, the Weyl system has to be modified. Instead of the group of translations, appearing naturally in the formula 
giving $W(\xi)$ (see (\ref{wey})), one has to work with the magnetic translations, forming a sort of generalized projective representation of 
$\mathbb R^N$. This is presented in the second Section.

In the third Section, the formula $\Op^A$ for the functional calculus with magnetic field is deduced. For this we apply the same strategy as in 
Section 1, but using now the magnetic Weyl system, introduced in Section 2. The setting relies on the choice of a vector potential, but now gauge 
covariance is available; equivalent vector potentials lead to unitarily equivalent operators. The basic expression for $\Op^A(f)$ requires rather 
strong conditions on the function $f$. But for large classes of magnetic fields one can extend it, as in the non-magnetic case, to all tempered 
distributions, by a suitable interpretation of $\Op^A(f)$ as a linear continuous operator from the Schwartz space $\mathcal S(\mathbb{R}^N)$ to 
its dual. This is based on a study of the distribution kernel of this operator. These and some other fundamental facts are also considered in 
Section 3.

Beyond these operators lies a symbolic calculus which is {\it manifestly gauge invariant}, being defined only in terms of the magnetic field. The 
composition is a magnetic correction of the Moyal product, while the involution is just the usual complex conjugation of functions. In Section 4 we 
study this symbolic calculus. Once again an important problem is to extend the formulae when obvious integrability conditions are not satisfied. 
This is a more complicated task for the product than it was for the quantization itself, especially if one aims at obtaining $^*$-algebras. We 
postpone the application of the machinery of oscillatory integrals and classical symbol function spaces to a future article. For our present 
purposes the strategy of extension by duality methods (see \cite{Gracia1}, \cite{Gracia2} for the non-magnetic case) is more fruitful. It will lead 
to a large, interesting $^*$-algebra of distributions which will be called {\it the magnetic Moyal algebra}.

Of course, for certain problems, a well-justified norm on (restricted) $^*$-algebras of symbols could be very useful. It happens that this is 
easier to achieve after performing a partial Fourier transform. Surprisingly, one naturally encounters certain $C^*$-algebras which were studied 
in pure mathematics, with little connection with physics. These are special types of twisted crossed products, associated to twisted actions of 
$\mathbb R^N$ on suitable abelian $C^*$-algebras of position observables. They were already related to quantum magnetic fields in \cite{Purice}; 
see also \cite{Georgescu}, \cite{Bellissard1} and \cite{Bellissard2} for related works. In a future publication we shall extend their study and 
outline the connection with our pseudodifferential calculus.

Concerning the difference between the expressions (\ref{ei}) (or (0.3)) and (\ref{noi}) some comments are necessary. In Subsection 3.4 we shall 
outline some situations when they give the same result. By admitting the convenient assumption that the components of $A$ are smooth 
functions with tempered growth, both (\ref{ei}) and (\ref{noi}) can be extended to any tempered distribution $f$. Then {\it we shall have 
$\Op_A(f)=\Op^A(f)$ for all $f$ if and only if $A$ is linear} (this is one of the most important cases appearing usually in the literature). Remark 
that this condition corresponds to a constant magnetic field, but {\it it is not gauge invariant}: for some other $A'$ with $dA'=B=\text{const}$, 
$\Op_{A'}$ and $\Op^{A'}$ will be different! (The pseudodifferential calculus corresponding to a linear $A$ was developped in \cite{Boutet} in 
connection with some problems in pure PDE theory; in fact the term ``magnetic field'' is never explicitly mentioned.) One will also have 
$\Op_A(f)=\Op^A(f)$ {\it for any} $A$ if \textit{f} is a polynomial of order $\le 2$. The most studied magnetic operators are the Schr\"odinger 
magnetic Hamiltonians $(P-A(Q))^2+V(Q)$, cf \cite{Avron} and \cite{Raikov} for instance. They are obtained by quantizing the sum between a 
quadratic function dependind only on $p$ and a function depending only on $x$, so no care is needed in this case. However, even their study may 
involve applying the Weyl calculus to more complicated symbols. Anyhow, for polynomials of order three in $p$, (\ref{ei}) and (\ref{noi}) already 
give different results. We remark that using the Weyl calculus coupled with the minimal coupling principle (as in (\ref{ei})) for a non-constant 
magnetic field and a complicated symbol is still legitimate as long as this is a technical tool and not the quantization of the classical observable 
represented by the symbol. This is often the case in solid state physics, in arguments concerning the Peierls substitution, as in \cite{Gerard} and 
\cite{Teufel}.

Our feeling is that the magnetic Weyl calculus elaborated in this paper is both an interesting mathematical object and a significant formalism for 
theoretical physics. 
From the mathematical point of view, we consider interesting to use our calculus for some specific classes of symbols and obtain more detailed results and also to connect it with strict deformation quantization. Moreover, obtaining precise estimations on the $C^*$-norm of the objects in the Moyal algebra and some variants of Calderon-Vaillancourt theorems is of much interest in spectral analysis, deformation quantization and semiclassical limit. These results may be then applied for quantum Hamiltonians (of Schr\"{o}dinger or relativistic type) and obtain spectral and propagation information and to study their semiclassical limit and its dependence on the chosen quantization procedure.
In this paper we intended to be accessible to people that are only vaguely familiar with pseudodifferential theory; more technical 
developments or applications are defered to future works. We were encouraged in this attitude by a discussion with Joseph Avron and Omri Gat. A paper devoted to some $C^*$-algebras aspects of our magnetic Weyl calculus is in preparation in collaboration with Serge Richard. 

\section{Canonical commutation relations and pseudodifferential calculus without magnetic fields}

One of the main virtues of the standard pseudodifferential calculus (in Weyl form) lies in the fact that it gives an answer to a fundamental 
problem in Quantum Mechanics. It can be figured out as the quantization of a physical system composed of a non-relativistic particle without 
internal structure, moving in an Euclidean configuration space. We review this topic briefly, since it gives a solid motivation for our later 
treatment of the case in which a magnetic field is added. Further details may be found in \cite{Folland} and \cite{Hormander1} for example. At the 
root of this approach lie the canonical commutation relations satisfied by the basic observables of the system, the positions and the momenta, 
and this is the main point we want to emphasize.	

\subsection{Framework}

We have in view an $N$-dimentional non-relativistic particle without internal structure (called simply {\it a particle}) that is described classically 
in {\it the phase space} $\Xi:=X\times X^{\star}$, where $X:=\mathbb{R}^N$ is {\it the configuration space} and $X^{\star}$ is its dual. The space 
$\Xi$ is naturally endowed with the symplectic form $\sigma:\Xi\times\Xi\rightarrow\mathbb{R}$ given by 

$$
\sigma((q',p'),(q'',p'')):=q''\cdot p'-q'\cdot p'',
$$
where $q\cdot p$ denotes the canonical pairing on $X\times X^\star$.

The classical observables are (smooth) real functions defined on $\Xi$. A particular role is played by the Poisson bracket 

$$
\{f,g\}:=\sigma(\nabla f,\nabla g)=\sum_{j=1}^N \left(\partial_{p_j}f\;\partial_{q_j}g-\partial_{p_j}g\ \partial_{q_j}f\right).
$$
Real functions of class $C^\infty$ on the phase space form an infinite-dimentional Lie algebra under the pointwise vector operations and the 
Poisson bracket. 

The associated quantum system is described on the Hilbert space $\mathcal{H}=L^2(X)$ in terms of the family of self-adjoint operators 
$(Q_j)_{j=1,...,N}$ and $(P_j)_{j=1,...,N}$ (here $Q_j$ is the operator of multiplication with the j-th component of the variable in $\mathcal{H}$ 
and $P_j:=-i\partial_j$). The operators $(Q_j,P_j)_{j=1,\dots,N}$ are the quantum version of the classical observables position and momenta, 
given by the the canonical variables in phase space $q_1,\cdots,q_N,p_1,\dots,p_N$. These canonical variables satisfy the relations

$$
\{q_i,q_j\}=0,\ \{p_i,p_j\}=0,\ \{p_i,q_j\}=\delta_{ij}, \ \ i,j=1,\dots,N
$$
and the corresponding quantum observables should satisfy at their turn 

$$
i[Q_i,Q_j]=0,\ i[P_i,P_j]=0,\ i[P_i,Q_j]=\delta_{ij}, \ \ i,j=1,\dots,N,
$$
as they actually do, at least formally.

\subsection{The Weyl system}

In principle, the choice of the Hilbert space $L^2(X)$ and of the explicit form of the operators $Q_j$ and $P_j$ should be justified. It is widely 
accepted the vague prescription that to the canonical variables $q_j$ and $p_j$ one should ascribe self-adjoint operators $\mathfrak{Op}(q_j)$ 
and $\mathfrak{Op}(p_j)$ acting in some Hilbert space $\cal H$, satisfying

\begin{equation}
i[\mathfrak{Op}(q_i),\mathfrak{Op}(q_j)]=0,\ i[\mathfrak{Op}(p_i),\mathfrak{Op}(p_j)]=0,\ i[\mathfrak{Op}(p_i),\mathfrak{Op}(q_j)]=\delta_{ij}, \ \ 
i,j=1,\dots,N.
\end{equation}
But an axiomatic approach relying on this formulae is hard to conceive. The typical difficulties related to the (inevitable) non-boundedness of the 
operators $\Op(q_j)$ and $\Op(p_j)$ cannot be solved by a priori arguments.

For this and for several other reasons, it is preferable to rephrase all in term of bounded operators. For $q\in X$ and $p\in X^\star$, let us set 
$U(q):=e^{-iq\cdot P}$ and $V(p):=e^{-iQ\cdot p}$. These are unitary operators in $L^2(X)$ given explicitly by

\begin{equation}\label{SRep}
[U(q)u](y)=u(y-q)\ \ \text{and}\ \ [V(p)u](y)=e^{-iy\cdot p}u(y),\ \ u\in L^2(X),\ y\in X.
\end{equation}
The maps $q\mapsto U(q)$ and $p\mapsto V(p)$ are strongly continuous unitary representations of $X$, respectively $X^*$, in $L^2(X)$ and {\it 
the Weyl form of the canonical commutation relations} 

\begin{equation}\label{WCCR}
U(q)V(p)=e^{iq\cdot p}\;V(p)U(q),\ \ q\in X,\ p\in X^\star
\end{equation}
holds. Now there is no ambiguity in addressing the abstract problem of the classification of triples $(\mathcal H,U,V)$, where $\mathcal H$ is a 
Hilbert space and $U:X\rightarrow\mathcal U(\mathcal H)$, $V:X^\star\rightarrow\mathcal U(\mathcal H)$ are strongly continuous unitary 
representations satisfying (\ref{WCCR}). And there is a simple answer, given by the Stone-von Neumann Theorem (for a more explicit statement 
and for the proof we send to \cite{Folland}): If one also assumes irreducibility of the family $\{U(q),V(p)\mid q\in X,p\in X^\star\}$, then any 
solution is unitarily equivalent to (\ref{SRep}) (which is called {\it the Schr\"odinger representation}). And a non-irreducible triple is just a multiple 
of this Schr\"odinger representation.

A convenient way to condense the two objects $U$ and $V$ into a single one is to define {\it the Weyl system} $\{W(\xi)\mid 
\xi\in\Xi\}\subset\mathcal U(\mathcal H)$ by

\begin{equation}\label{wey}
W(q,p):=e^{\frac{i}{2}q\cdot p}\;U(-q)V(p)=e^{-\frac{i}{2}q\cdot p}\;V(p)U(-q),\ \ q\in X,\ p\in X^\star.
\end{equation}
A short calculation shows that $W$ satisfies

\begin{equation}
W(\xi)W(\eta)=e^{\frac{i}{2}\sigma(\xi,\eta)}\;W(\xi+\eta),\ \ \ \xi,\eta\in\Xi,
\end{equation}
i.e. $W$ is a projective representation of the group $\Xi$ with $2$-cocycle (phase factor) $e^{\frac{i}{2}\sigma}$.

Of course, $W$ can be defined for any abstract triple $(\mathcal H,U,V)$. But, as a consequence of the Stone-von Neumann Theorem, it is 
enough to work with the Schr\"odinger representation. The corresponding $W$ will be called {\it the Schr\"odinger Weyl system} and is explicitely 
given on $L^2(X)$ by 

\begin{equation}
[W(q,p)u](y)=e^{-i\left(\frac{1}{2}q+y\right)\cdot p}\;u(y+q).
\end{equation}
The representations $U$ and $V$ can be recovered easily from $W$ by $U(q)=W(-q,0)$ and $V(p)=W(0,p)$. One easily justifies the formula 
$W(\xi)=e^{-i\sigma(\xi,R)}$, where $R=(Q,P)$; $\sigma(\xi,R)$ signifies here the (suitable defined) self-adjoint operator $Q\cdot p-q\cdot P$.

The Weyl system is a convenient way to codify the commutation relations between the basic operators $Q$ and $P$. In the next paragraph, the 
quantization by pseudodifferential operators will be obtained as an integrated form of this Weyl system.

\subsection{Pseudodifferential operators}

If a family of self-adjoint operators $S_1,\dots,S_m$ is given such that for any $i,j$, $S_i$ and $S_j$ commute, then one can define a functional 
calculus for this family by one of the two formulae

$$
f(S)=\int_{\mathbb R^m}f(\lambda)dE_S(\lambda)=\int_{\mathbb R^m}dt\;\check{f}(t)e^{-it\cdot S}.
$$
Here $E_S$ is the spectral measure (on $\mathbb R^m$) of the family $S_1,\dots,S_m$, under suitable assumptions $\ t\cdot 
S:=t_1S_1+\dots+t_mS_m$ is a well-defined self-adjoint operator and $\check{f}$ is the inverse Fourier tranform of $f$, conveniently 
normalized. 

If, once again, $S_1,\dots,S_m$ are self-adjoint, but they no longer commute, there is usualy no reasonable spectral measure $E_S$. One can 
try to use the operator version of the Fourier inversion formula to define a functional calculus. The key point would be the ability of defining a 
suitable analogue of $e^{-it\cdot S}$. This strategy is outlined in \cite{Anderson1} (see also \cite{Anderson2}) for very general situations. But the 
properties of the resulting functional calculus are quite modest if the commutation relations of the operators $S_j$ have no interesting 
peculiarities.

We shall show how this program can be implemented for the case $m=2N$, $S_j=Q_j$ if $j=1,\dots,N$ and $S_j=P_j$ for $j=N+1,\dots,2N$. In an 
analoguous, but more complicated way, in Section 3 we shall do the same for $S_j=Q_j$ if $j=1,\dots,N$ and $S_j=\Pi^A_j$ for $j=N+1,\dots,2N$, 
with $\Pi_j^A=P_j-A_j(Q)$ the $j$'th component of the magnetic momentum defined by a vector potential $A$. But we stop for a moment to fix 
some conventions on Fourier transforms that will also be useful later on.

In fact we are faced with two problems: normalization and the choice of a good definition on the symplectic space. The Lebegue measures on 
$X$, $X^\star$ and $\Xi$ are not the most convenient Haar (= positive, translational invariant Borel) measures, since they lead to the appearance 
of spurious constants. Let us start with two arbitrary Haar measures $dx$ on $X$ and $dp$ on $X^*$. One defines at the level of tempered 
distributions

$$
\mathcal{F}_X,\overline{\mathcal{F}}_X:\mathcal{S}'(X)\rightarrow\mathcal{S}'(X^*), \ \ \ 
\mathcal{F}_{X^\star},\overline{\mathcal{F}}_{X^\star}:\mathcal{S}'(X^\star)\rightarrow\mathcal{S}'(X),
$$
uniquely determined by the following actions on integrable functions: 

$$
(\mathcal{F}_Xu)(p)=\int_Xdx\;e^{-ix\cdot p}u(x), \ \ \ (\overline{\mathcal{F}}_Xu)(p)=\int_Xdx\;e^{ix\cdot p}u(x),
$$

$$
(\mathcal{F}_{X^\star}v)(x)=\int_{X^\star}dp\;e^{-ix\cdot p}v(p), \ \ \ (\overline{\mathcal{F}}_{X^\star}v)(p)=\int_{X^\star}dp\;e^{ix\cdot p}v(p).
$$
It is easily shown that there exists $c>0$ such that 

$$
\overline{\mathcal{F}}_{X^*}\circ\mathcal{F}_X=c\;\text{id}_{\mathcal S'(X)}\ \ \ 
\text{and}\ \ \  \mathcal{F}_{X}\circ\overline{\mathcal{F}}_{X^\star}=c\;\text{id}_{\mathcal S'(X^\star)}.
$$
Thus, by redefining $dx$ and $dp$, one gets $\mathcal{F}_X^{-1}=\overline{\mathcal{F}}_{X^\star}$ and 
$\mathcal{F}_{X^\star}^{-1}=\overline{\mathcal{F}}_X$. We fix such a choice for $dx$ and $dp$, but obviously $d\xi:=dx\otimes dp$ does not 
depend on this choice. We also set the symplectic Fourier transforms

$$
\mathcal F_\Xi,\mathcal F_\Xi^{-1}:\mathcal S'(\Xi)\rightarrow\mathcal S'(\Xi),
$$
$$
(\mathcal F_\Xi f)(\xi):=\int_\Xi d\eta\;e^{-i\sigma(\xi,\eta)}f(\eta),\ \ \ (\mathcal F_\Xi^{-1} f)(\xi):=\int_\Xi d\eta\;e^{i\sigma(\xi,\eta)}f(\eta)
$$
and note that $\mathcal F_\Xi=\mathcal I\circ(\mathcal F_X\otimes\overline{\mathcal F}_{X^\star})$, where $\mathcal I:\mathcal S'(X^\star\times 
X)\rightarrow\mathcal S'(X\times X^\star)$, $(\mathcal I g)(x,p):=g(p,x)$.

Now, for any Weyl system $(\mathcal H,W)$ we define (at least) for functions $f:\Xi\rightarrow\mathbb C$ with integrable symplectic Fourier 
transform

\begin{equation}
\Op(f):=\int_\Xi d\xi\;(\mathcal F_\Xi^{-1} f)(\xi)\;W(\xi).
\end{equation}
We do not insist on the precise interpretation of this formula; this will be done later on in the more complicated, magnetic case.

Once again, by the Stone-von Neumann Theorem, we are satisfied with the case of the Schr\"odinger representation. By introducing the explicit 
form of the Schr\"odinger Weyl system, one gets immediatly for any $u\in L^2(X)$

\begin{equation}
[\Op(f)u](x)=\int_Xdy\int_{X^\star}dp\;e^{i(x-y)\cdot p}f\left(\frac{x+y}{2},p\right)u(y)
\end{equation}
and this is exactly the Weyl prescription to quantize classical symbols. 

We note that $\Op(f)$ is an integral operator with kernel $K_f(x,y):=[(1\otimes\overline{\mathcal F}_{X^\star})f](\frac{x+y}{2},x-y)$. Then, by 
an elementary application of Schwartz's Kernel Theorem, one gives a sense to $\Op(f)$ for any $f\in\mathcal S'(\Xi)$ as a continuous linear 
operator from $\mathcal S(X)$ to $\mathcal S'(X)$. In fact all these operators are of the form $\Op(f)$ for some unique tempered distribution 
$f$. 

\subsection{The Moyal algebra}

We turn now to the symbolic calculus. It is easy to see that by setting

\begin{equation}\label{Moya}
(f\circ g)(\xi):=4^N\int_\Xi d\eta\int_\Xi d\zeta\;e^{-2i\sigma(\xi-\eta,\xi-\zeta)}f(\eta)g(\zeta)
\end{equation}
one will have $\Op(f)\Op(g)=\Op(f\circ g)$, and that $\Op(f)^*=\Op(f^\circ)$, with $f^\circ(x):=\overline{f(x)}$.

The non-commutative composition law $\circ$ is often called {\it the Moyal product} (or {\it the Weyl product}). It makes sense for suitable 
symbols, say $f,g\in\mathcal S(\Xi)$. For many purposes it is useful  to extend it to larger classes of functions and distributions. The standard 
approach (see \cite{Folland}, \cite{Hormander1}, \cite{Hormander2}, \cite{Shubin} and many others) is via oscillatory integrals. Better suited to 
our setting is the approach by duality of \cite{Antonets}, \cite{Gracia1} and \cite{Gracia2} that we review now briefly. 

Let us denote by $(\cdot,\cdot)$ the duality $\mathcal{S}'(X)\times\mathcal{S}(X)\rightarrow\mathbb{C}$. By a simple calculation we see that 
for any three functions $f$, $g$ and $h$ in $\mathcal{S}(\Xi)$ we have
$$
(f,g\circ h)=(f\circ g,h)=(h,f\circ g)=(h\circ f,g)=(g,h\circ f).
$$
Thus, we can extend $\circ$ to mappings $\mathcal S(\Xi)\times\mathcal S'(\Xi)\rightarrow\mathcal S'(\Xi)$ and $\mathcal S'(\Xi)\times\mathcal 
S(\Xi)\rightarrow\mathcal S'(\Xi)$ by $(f\circ G,h):=(G,h\circ f)$ and $(F\circ g,h):=(F,g\circ h)$, for $f,g,h\in\mathcal S(\Xi)$ and 
$F,G\in\mathcal S'(\Xi)$. This is already useful and allows composing $n$ symbols if all except one are in the Schwartz space.

Now set $\mathcal M(\Xi):=\{F\in\mathcal S'(\Xi)\mid F\circ\mathcal S(\Xi)\subset\mathcal S(\Xi)\; \text{ \sl and}\ \;\mathcal S(\Xi)\circ 
F\subset\mathcal S(\Xi)\}$. Just by some abstract nonsense one checks that $\mathcal M(\Xi)$ is a $^*$-algebra under the (extension of) the 
Moyal product $\circ$ and the involution $^\circ$. In \cite{Gracia1} $\mathcal M(\Xi)$ is called {\it the Moyal algebra} and some of its properties 
are studied. In particular it is shown that $\mathcal M(\Xi)$ is stable under all sort of Fourier transforms, it contains all the distributions with 
compact support and (thus) large classes of analytic functions. It also contains the family of $C^\infty$ functions on $\Xi$ with all the 
derivatives dominated by the same (arbitrary) polynomial.

We will reconsider this topic in greater detail in Section 4, where the magnetic field will also be present.

\section{The magnetic Weyl system}

We consider a quantum particle without internal structure moving in $X=\mathbb{R}^N$, in the presence of a variable magnetic field. The {\it 
magnetic field} is described by a closed continuous field of 2-forms $B$ defined on $\mathbb{R}^N$. In the standard coordinate system on 
$\mathbb{R}^N$, it is represented by a continuous function taking real antisymetric matrix values and verifying the cocycle relation 
$\partial_jB_{kl}+\partial_kB_{lj}+\partial_lB_{jk}=0$ in a distributional sense. The reader will verify for himself that many constructions and 
assertions will still be valid for locally integrable fields; we assumed continuity for simplicity and to have a uniform framework.

It is well-known that any such field $B$ may be written as the (distributional) differential $dA$ of a field of 1-forms $A$, {\it the vector 
potential}, that is highly non-unique (the gauge ambiguity); by using coordinates, one has $B_{jk}=\partial_jA_k-\partial_kA_j$ for each 
$j,k=1,\cdots,N$.

In this Section we shall deduce a formula for the analog of the Weyl system of Subsection 1.2, but in which the magnetic field is also taken into 
account. One may proceed as in Subsection 1.2, with the single modification which consists in  replacing the translations by the magnetic 
translations (exponentials of $q\cdot\Pi^A$, where $\Pi^A:=P-A(Q)$ is the magnetic momentum). Just for a change, we proceed in a different, 
but equivalent, way. First we get directly the formula for our magnetic Weyl system by exponentiating the self-adjoint operators 
$\sigma[(q,p),(Q,\Pi^A)]=Q\cdot p-q\cdot(P-A(Q))$, $(q,p)\in\Xi$. Then the magnetic translations and the magnetic form of the Weyl 
commutation relations are deduced as consequences. 

The magnetic translations have appeared since long in the physical literature (see \cite{Luttinger} and \cite{Zak} for example), especially in 
connection with problems in solid state physics. Most of the times they were used for the case of a constant field; some references are 
\cite{Bellissard1}, \cite{Bellissard2}, \cite{Helffer1}, \cite{Helffer2} and \cite{Nenciu}.

We stress that the new objects appearing in the magnetic case are two phase factors: One is defined as the imaginary exponential of the 
circulation of the vector potential; it enters the definition of the magnetic translations, the magnetic Weyl system and (as a consequence, in 
Section 3) in the expression of the magnetic pseudodifferential operators. The other one, an imaginary exponential of the flux of the magnetic 
field, appears in connection with multiplicative properties of the magnetic translations and of the magnetic Weyl system and (as a consequence, 
in Section 4) in the expression of the composition law defining the symbolic calculus. We hope that our treatment will constitute a source of 
unification of the various ``non-integrable phase factors'' scattered in the literature on quantum magnetic fields.

\subsection{The magnetic Weyl system}

Given a $k$-form $C$ on $X$ and a compact $k$-surface $\gamma\subset X$, we define
$$
\Gamma^C(\gamma):=\int_\gamma C
$$
(this integral having a well-defined invariant meaning). We shall mainly encounter circulations of 1-forms along linear segments 
($\gamma=[x,y]$) and fluxes of 2-forms through triangles ($\gamma=<x,y,z>$).

We denote by $\mathcal H$ the Hilbert space $L^2(X)$. For each $t\in\mathbb R$ we define
\begin{equation}\label{W-def}
W_t^A:\Xi\rightarrow\mathcal{U}(\mathcal{H}),\qquad W_t^A(x,p):=e^{-it(Q+tx/2)\cdot p}\Lambda^A(Q;tx)e^{itx\cdot P},
\end{equation}
where we introduced the exponential of the circulation of the vector potential
\begin{equation}\label{Phase-def}
\Lambda^A(q;x):=e^{-i\Gamma^A([q,q+x])}=e^{-ix\cdot\int_0^1ds\;A(q+sx)}.
\end{equation}

We make the convention that the vector potential will always be taken continuous. This is, indeed, always possible, since $B$ is supposed 
continuous, by {\it the transversal gauge}

\begin{equation}\label{transversal}
A_i(x)=-\sum_{j=1}^N\int_0^1ds\;B_{ij}(sx)sx_j.
\end{equation}
Non-continuous vector potentials are not really useful in our framework, but they could also be handled either directly or by exploiting gauge 
covariance.

The next Lemma says that $\{W_t^A(x,p)\}_{t\in \mathbb R}$ is the evolution group of the self-adjoint operator $Q\cdot p-x\cdot\Pi^A$, suitably 
defined.

\begin{lemma}
We have $W_t^A(x,p)=e^{-it\sigma[(x,p),(Q,\Pi^A)]}$, where the self-adjoint operator $\sigma[(x,p),(Q,\Pi^A)]$ is the closure of the restriction 
at $\;\mathcal S(X)$ of the sum $S+T$, with $S=Q\cdot p+x\cdot A(Q)$ and $T=-x\cdot P$. 
\end{lemma}

\begin{proof}
It is known that $S+T$ is indeed essentially self-adjoint on $C_c^\infty(X)$ (see \cite{Leinfelder-Simader}, \cite{CFKS}). Thus, we can apply Trotter's formula (see \cite{Reed}, Th. VII.31). 
We set $S=a(Q)$ and calculate

\begin{align*}
&\left(e^{-\frac{i}{n}ta(Q)}e^{\frac{i}{n}tx\cdot P}\right)^n\ =\ e^{-\frac{i}{n}ta(Q)}e^{\frac{i}{n}tx\cdot 
P}e^{-\frac{i}{n}ta(Q)}e^{-\frac{i}{n}tx\cdot P}\cdot\\\cdot\ e^{\frac{2i}{n}tx\cdot P}&e^{-\frac{i}{n}ta(Q)}e^{-\frac{2i}{n}tx\cdot 
P}e^{\frac{3i}{n}tx\cdot P}\dots e^{\frac{(n-1)i}{n}tx\cdot P}e^{-\frac{i}{n}ta(Q)}e^{-\frac{(n-1)i}{n}tx\cdot P}e^{\frac{ni}{n}tx\cdot 
P}=\\&=e^{-\frac{it}{n}\left[a(Q)+a\left(Q+\frac{tx}{n}\right)+\dots+a\left(Q+\frac{(n-1)tx}{n}\right)\right]}e^{itx\cdot P}.
\end{align*}
One notes the appearance of a Riemman sum at the exponent, hence the last expression converges strongly to $e^{-i\int_0^t 
ds\;a(Q+sx)}e^{itx\cdot P}$. The proof is ended by remarking that 
$$
\int_0^tds\;\{(y+sx)\cdot p+x\cdot A(y+sx)\}=ty\cdot p+\frac{t^2}{2}x\cdot p+\Gamma^A[y,y+tx].
$$
\end{proof}

Another, more annoying, proof would consist in showing that for all $\xi\in \Xi$, $\,t\mapsto W_t(\xi)$ is a strongly continuous unitary group in 
$\mathcal H$ and then doing the necessary derivations.

We note the obvious formula $W_t^A(\xi)=W_1^A(t\xi)$. The operator $W_1^A(\xi)$ will be denoted simply by $W^A(\xi)$. 

\begin{definition}
The family $\{W^A(\xi)\}_{\xi\in\Xi}$ will be called {\it the magnetic Weyl system associated to the vector potential $A$}. We write down here, for 
further use, the action of $W^A(\xi)$ on vectors $u\in\mathcal H=L^2(X)$:

\begin{equation}
\left[W^A(x,p)u\right](y)=e^{-i(y+x/2)\cdot p}e^{-i\Gamma^A([y,y+x])}u(y+x).
\end{equation}

\end{definition}

The usual Weyl system was a projective representation of $\Xi$. Now the situation is of the same nature, but more involved. For $x,y,q\in X$, let 
us define

\begin{equation}\label{omega}
\Omega^B(q;x,y):=e^{-i\Gamma^B(<q,q+x,q+x+y>)}.
\end{equation}
We note that this is a continuous function of $q$ for fixed $x$ and $y$, thus it defines a multiplication operator in $\mathcal H$.

\begin{proposition}\label{viatanoua}
For any $\ \xi=(x,k),\eta=(y,l)\in\Xi$ one has

\begin{equation}\label{formula}
W^A(\xi)W^A(\eta)=e^{\frac{i}{2}\sigma(\xi,\eta)}\Omega^B(Q;x,y)W^A(\xi+\eta).
\end{equation}
\end{proposition}

\begin{proof}
By Stokes Theorem coupled with the relation $B=dA$, one gets for any $x,y,q\in X$ the equality 
$\Omega^B(q;x,y)=\Lambda^A(q;x)\Lambda^A(q+x;y)\left[\Lambda^A(q;x+y)\right]^{-1}$. Then (\ref{formula}) follows by a routine calculation.
\end{proof}

Let us denote by $C(X;U(1))$ the group (with pointwise multiplication) of all continuous functions on $X$, taking values in $U(1)$, the 
multiplicative group of complex numbers of modulus 1. One can interpret $\Omega^B$ as a function $\Omega^B:X\times X\rightarrow C(X;U(1))$. 
This function satisfies the following 2-cocycle conditions:

\begin{equation}\label{trebuie}
\begin{array}{l}
\Omega^B(q;x,0)=\Omega^B(q;0,y)=1,\\
\Omega^B(q;x+y,z)\Omega^B(q;x,y)=\Omega^B(q+x;y,z)\Omega^B(q;x,y+z).
\end{array}
\end{equation}
They follow easily by direct calculations (for the second one use Stokes Theorem for the closed 2-form $B$ and the tetrahedron of vertices 
$q,q+x,q+x+y$ and $q+x+y+z$), but are also easy consequences of Proposition \ref{viatanoua}. We also note that $\Omega^B(q;x,-x)=1$.

\subsection{The magnetic canonical commutation relations}

By restricting to $X$, respectively $X^\star$, we recover the usual {\it magnetic translations}, respectively the unitary group generated by the 
position operators:
\begin{equation}\label{aia}
\begin{array}{l}
U^A(x):=W^A(-x,0)=\Lambda^A(Q;-x)e^{-ix\cdot P}=\Lambda^A(Q;-x)U(x),\\
V(p):=W^A(0,p)=e^{-iQ\cdot p}.\\
\end{array}
\end{equation}
One has, analogously to (\ref{wey}), 
\begin{equation}\label{label}
W^A(x,p):=e^{\frac{i}{2}x\cdot p}\;U^A(-x)V(p)=e^{-\frac{i}{2}x\cdot p}\;V(p)U^A(-x),\ \ x\in X,\ p\in X^\star.
\end{equation}
We get easily from (\ref{formula}) (or by direct calculation) the commutation rules

\begin{equation}\label{una}
V(p)V(k)=V(k)V(p),\ \ U^A(x)V(p)=e^{ix\cdot p}V(p)U^A(x)
\end{equation}
and
\begin{equation}\label{duo}
U^A(x)U^A(y)=\Omega^B(Q;-x,-y)U^A(x+y),
\end{equation}
that are the magnetic extension of the Weyl form of the canonical commutation relations.

For any $x\in X$ and any $p\in X^\star$, the applications $\mathbb{R}\ni t\mapsto U^A(tx)\in\mathcal{U}(\mathcal{H})$ and $\mathbb{R}\ni 
t\mapsto V(tp)\in \mathcal{U}(\mathcal{H})$ are 1-parameter unitary groups on $\mathcal{H}$. We define self-adjoint generators (chosing 
$x=e_j$, resp. $p=\epsilon_j$ the j-th element of the canonical orthogonal basis in $\mathbb{R}^N$)

\begin{equation}
\begin{array}{l}
Q_j:=i\left.\frac{\partial}{\partial t}\right|_{t=0}V(t\epsilon_j),\\
\Pi^A_j:=i\left.\frac{\partial}{\partial t}\right|_{t=0}U^A(te_j)=P_j-A_j(Q).
\end{array}
\end{equation}
On the common domain formed of $C^\infty$-functions with compact support we have the following commutation relations:
\begin{equation}
i[Q_j,Q_k]=0,\ \ i[Q_j,\Pi_k]=1,\ \ i[\Pi_j,\Pi_k]=B_{jk}(Q).
\end{equation}
If the magnetic field is not constant (or at least polynomial) they are much more complex than in the non-magnetic case; the successive 
commutators of the components of $B$ with the magnetic momenta are non-trivial.

\section{Magnetic pseudodifferential operators}

Our intention is to elaborate a functional calculus for the {\it non-commutative} family of self-adjoint operators $\{Q_j,\Pi_k\}_{j,k=1}^N$. We 
shall call it {\it the Weyl calculus with magnetic field}. As in the non-magnetic case, we obtain it by an analog of the Fourier inversion formula, the 
magnetic Weyl system of the preceding Section playing the part of the imaginary exponential. The resulting formula has the right gauge 
covariance. The operators involved are all integral operators and by the Kernel Theorem they can also be defined for symbols which are 
tempered distributions. The problem of identifying finite-rank, Hilbert-Schmidt and compact operators is also addressed. For this, an extension of 
the classical Fourier-Wigner transform (cf. \cite{Folland}) is of great help. It also shows a posteriori the irreducibility of our magnetic Weyl 
system. In the final part of the Section we compare the magnetic Weyl calculus and the composition of the usual Weyl caculus with the minimal 
coupling prescription. They are different but, striking enough, they give the same result in many important cases. This explains perhaps the fact 
that the quantization of observables in a magnetic field has not been treated properly before in a suitable generality. We note, however, that in 
\cite{Luttinger} one finds (in a non-systematic setting) the right attitude for the case of periodic symbols depending only on $p$. We thank 
Professor Nenciu for drawing our attention to this reference.

Let us finally remark that for a constant magnetic field, using a linear vector potential, one is lead to a change of the canonic symplectic form of 
the space $\Xi=X\times X^{\star}$, while for non-constant magnetic fields one is quantizing a symplectic manifold (with a 
non-constant symplectic form of the special type $\sigma_B:=\sigma+B$) associated to the same linear space $\Xi$.

\subsection{The functional calculus}

We define the linear mapping
\begin{equation}\label{Op-def}
\mathfrak{Op}^A:\mathcal{F}_\Xi L^1(\Xi)\rightarrow\mathcal{B}(L^2(X)),\qquad \mathfrak{Op}^A(f):=\int_\Xi d\xi\;(\mathcal{F}_\Xi^{-1} 
f)(\xi)\;W^A(\xi)
\end{equation}
in weak sense: if $u,v\in\mathcal H:=L^2(X)$, then $\left<v,\Op^A(f)u\right>=\int_\Xi d\xi\;(\mathcal{F}_\Xi^{-1} f)(\xi)\left<v,W^A(\xi)u\right>$.
It clearly satisfies the estimate $\|\mathfrak{Op}^A(f)\|\leq\|\mathcal{F}_\Xi^{-1} f\|_{L^1}$. 

Using the expression of the operators $W^A(\xi)$ given in (\ref{W-def}) and (\ref{Phase-def}) we obtain, at least formally,  the explicit form of 
the operators $\mathfrak{Op}^A(f)$:

\begin{equation}\label{Op-form}
\left(\mathfrak{Op}^A(f)u\right)(x)=\int_Xdy\int_{X^\star}dk\;e^{i(x-y)\cdot k}\tilde{\Lambda}^A(x,y)f\left(\frac{x+y}{2},k\right)u(y),
\end{equation}
where $\tilde{\Lambda}^A(x,y):=e^{-i\Gamma^A([x,y])}=\Lambda^A(x;y-x)$. 
For $A=0$ this is the usual Weyl prescription to quantize a classical symbol, encountered in the theory of pseudodifferential operators. For general 
(continuous) $A$ this is, in our opinion, the right formula that should stand for $f(Q,\Pi^A)$. 

In fact, the precise sense of (\ref{Op-def}) and (\ref{Op-form}) and of their equivalence depend on our assumptions on $f$ and $u$. In the next paragraphs, under certain 
hypothesis on the magnetic field, we shall cover the very general case in which $f$ is a tempered distribution; then both formulae will make sense with a suitable 
reinterpretation and actually define the same object. If $f$ is subject to suitable strong 
decay assumptions, then no special condition is needed (except our standing convention that $A$ is continuous). All is smooth, for example, if $f$ 
is in the Schwartz class $\mathcal S(\Xi)$. On the other hand, once again without any assumption on the magnetic field, (\ref{Op-def}) can be extended 
straightforwardly to $f$'s that are Fourier transforms of bounded complex measures on $\Xi$. Now, of course, (\ref{Op-form}) needs a reinterpretation.

To advocate our choice of the mapping $\Op^A$, an important point is to note {\it gauge covariance}: 

\begin{proposition}\label{gaco} Let $A$ and $A'$ be two continuous vector potentials defining the same continuous magnetic field: $dA=B=dA'$. 
Then there exists a real $C^1$-function $\rho$ on $X$ such that $A'=A+\nabla\rho$ and we have $\ 
e^{i\rho(Q)}W^A(\xi)e^{-i\rho(Q)}=W^{A+\nabla\rho}(\xi)$ for all $\xi\in\Xi$ and 
$e^{i\rho(Q)}\mathfrak{Op}^A(f)e^{-i\rho(Q)}=\mathfrak{Op}^{A+\nabla\rho}(f)$ for all $f\in\mathfrak{F}_\Xi L^1(\Xi)$. 
\end{proposition}

\begin{proof}
It is well-known (cf. \cite{Leinfelder} for example) that if $dA=B=dA'$ and $A,A'$ have $L^1_{\text{loc}}$-components, then there exists $\rho$ 
(in some suitable local Sobolev space that does not matter here) such that $A'-A=\nabla\rho\;$ in distributional sense. Now, since in our case $A$ 
and $A'$ are continuous, $\rho$ will be of class $C^1$ by a simple argument. The two identities are verified by trivial calculations based on the 
relation
$$
e^{i\rho(Q)}e^{ix\cdot P}e^{-i\rho(Q)}=e^{-i[\rho(Q+x)-\rho(Q)]}e^{ix\cdot P}=e^{-ix\cdot\int_0^1ds\;\nabla\rho(Q+sx)}e^{ix\cdot P}.
$$
\end{proof}

{\bf Remark.} One implements Planck's constant at the level of the physical momentum, by setting $P=\hbar D:=-i\hbar\nabla$. This gives for the 
magnetic Weyl system

$$
W_\hbar^A(x,p)=e^{-i\left(Q+\frac{\hbar}{2}x\right)\cdot p}e^{-\frac{i}{\hbar}\Gamma^A([Q,Q+\hbar x])}e^{i\hbar x\cdot D}
$$
and the $\hbar$-dependent magnetic $2$-cocycle will be $\Omega_\hbar^B(q;x,y)=e^{-\frac{i}{\hbar}\Gamma^B(<q,q+\hbar x,q+\hbar x+\hbar 
y>)}$. We collect here, for the convenience of the reader, formulae for the magnetic Weyl calculus

$$
\left(\mathfrak{Op}_\hbar^A(f)u\right)(x)=\hbar^{-N}\int_Xdy\int_{X^\star}dk\;e^{\frac{i}{\hbar}(x-y)\cdot 
k}e^{-\frac{i}{\hbar}\Gamma^A([x,y])}f\left(\frac{x+y}{2},k\right)u(y)
$$
and for the magnetic Moyal product (subject of Section 4)

$$
\left(f\circ_\hbar^B g\right)(\xi)=\left(\frac{2}{\hbar}\right)^{2N}\int_\Xi d\eta \int_\Xi d\zeta \; 
e^{-2\frac{i}{\hbar}\sigma(\xi-\eta,\xi-\zeta)}e^{-\frac{i}{\hbar}\Gamma^B(<q-y+x,x-q+y,y-x+q>)}f(\eta)g(\zeta).
$$

In the sequel $\hbar$ will always be $1$.

{\bf Remark.} One often uses instead of (\ref{prima}) {\it the $\tau$-quantizations} ($\tau\in[0,1]$), given by (cf. \cite{Shubin})
$$
[\mathfrak{Op}_{(\tau)}(f)u](x):=\int_{\mathbb R^{2N}}dy\;dp\;e^{i(x-y)\cdot p}f\left((1-\tau)x+\tau y,p\right)u(y).
$$
They are somehow connected with the ordering of $Q$ and $P$ in the expression of $f(Q,P)$. The cases $\tau=0$ and $\tau=1$ are called 
respectively {\it the right} and {\it the left quantization}. Rather often, in texbooks, only the case $\tau=0$ is treated. But the Weyl prescription 
$\Op\equiv\Op_{(1/2)}$ is preferred in Quantum Mechanics because of its nice property $\Op(f)^*=\Op(\overline f)$.

We obtain the magnetic analog of $\Op_{(\tau)}$ by replacing $W^A(\xi)$ with

$$
W^A_{(\tau)}(x,p):=e^{i(1-\tau)x\cdot p}U^A(-x)V(p)=e^{-i\tau x\cdot p}V(p)U^A(-x),\ \ x\in X,p\in X^\star.
$$
A short formal calculation shows that the definition $\;\Op^A_{(\tau)}(f):=\int_\Xi d\xi\;\left(\mathcal F_\Xi^{-1}f\right)(\xi)\;W^A_{(\tau)}(\xi)\;$ 
leads to
$$
[\mathfrak{Op}^A_{(\tau)}(f)u](x):=\int_{\mathbb R^{2N}}dy\;dp\;e^{i(x-y)\cdot p}e^{-i\Gamma^A([x,y])}f\left((1-\tau)x+\tau y,p\right)u(y),
$$
which allows a rigorous treatment analog to that given for $\Op^A\equiv\Op^A_{(1/2)}$ in the sequel.

\subsection{The distribution kernel}

$\mathfrak{Op}^A(f)$ is an integral operator having a kernel that can be defined in terms of $f$ and ``the phase function'' $\tilde{\Lambda}^A$. In 
fact let us introduce the one-to-one linear change of variables $(x,y)\mapsto S(x,y):=\left(x+\frac{y}{2},x-\frac{y}{2}\right)$ and denote by the 
same symbol $S$ the induced transformation on functions $(S\Phi)(x,y):=\Phi(S(x,y))=\Phi(x+y/2,x-y/2)$. The explicit form of the inverse is 
$S^{-1}(x,y)=\left(\frac{x+y}{2},x-y\right)$. We can now define (on $\mathcal S(\Xi)$ for instance) the map 

\begin{equation}\label{KA-def}
K^A:=\tilde{\Lambda}^AS^{-1}(\mathbf{1}\otimes\overline{\mathcal{F}}_{X^\star}),
\end{equation}
composed of a partial Fourier transform, a change of variables and a multiplication operator. It is easy to verify that $\mathfrak{Op}^A(f)$ is the 
integral operator with kernel $K^Af$. For functions $\Phi$ defined on $X\times X$ we shall denote by $\mathfrak{Int}(\Phi)$ the integral operator 
on $L^2(X)$ with kernel $\Phi$, so that one can write $\mathfrak{Op}^A(f)=\mathfrak{Int}(K^Af)$.  

For further use we shall introduce two more notations, trying to emphasize the special role played by the phase factor $\tilde{\Lambda}^A$. We 
define ``the zero magnetic field analog'' of $K^A$, the map $K:=S^{-1}(\mathbf{1}\otimes\overline{\mathcal{F}}_{X^\star})$ and the magnetic 
integral operator associated to a kernel $\Phi$ as $\mathfrak{Int}^A(\Phi):=\mathfrak{Int}(\tilde{\Lambda}^A\Phi)$. With these notations one may write 
for any $f\in\mathcal S(\Xi)$

\begin{equation}\label{op-int}
\mathfrak{Op}^A(f)=\mathfrak{Int}(K^Af)=\mathfrak{Int}^A(Kf).
\end{equation}

We use now these facts to extend the operation $\Op^A$ to distributions. Let us assume that the components of the magnetic field are 
$C_{\text{pol}}^\infty$ functions, i.e. they are indefinitely derivable and any derivative is polynomially bounded. These type of functions are also 
called {\it with tempered growth}; their main virtue is that by multiplication they leave the Schwartz space $\mathcal S$ invariant, hence they 
define by duality multiplication operators on $\mathcal S'$. The formula (\ref{transversal}) for the transversal gauge shows that the vector 
potential $A$ can also be chosen of class $C_{\text{pol}}^\infty$. By easy calculations, $\tilde{\Lambda}^A$ will also be $C_{\text{pol}}^\infty$ in 
both variables. Then it is clear that $K^A$ defines isomorphisms $\mathcal{S}(\Xi)\overset{\sim}{\rightarrow}\mathcal{S}(X\times X)$ and 
$\mathcal{S'}(\Xi)\overset{\sim}{\rightarrow}\mathcal{S'}(X\times X)$. 

On the other hand, let us recall that for any finite dimensional vector space $\mathcal{V}$, the spaces $\mathcal{S}(\mathcal{V})$ and 
$\mathcal{S}'(\mathcal{V})$ are nuclear and we have linear topological isomorphisms (see for example \cite{Treves} Theorem 51.6 and its 
Corollary)

\begin{equation}\label{Nucl-1}
\mathcal{S}(X)\otimes\mathcal{S}(X)\cong\mathcal{S}(X\times X),\qquad \mathcal{S}'(X)\otimes\mathcal{S}'(X)\cong\mathcal{S}'(X\times X).
\end{equation}
Here the tensor product is the closure of the algebraic tensor product for the injective or the projective topologies, that coincide in this case (we 
refer to \cite{Treves} Theorem 50.1). We shall be interested in the following spaces of linear continuous operators: 
$\mathcal{L}[\mathcal{S}(X),\mathcal{S}'(X)]$, $\mathcal{L}[\mathcal{S}'(X),\mathcal{S}(X)]$ and 
$\mathcal{L}[\mathcal{S}(X)]\cong\mathcal{L}[\mathcal{S}'(X)]$. On all these spaces we consider the topology of uniform convergence on 
bounded sets. It is easy to see that we have the continuous linear injections

\begin{equation}\label{oper-ext}
\mathcal{L}[\mathcal{S}'(X),\mathcal{S}(X)]\subset\mathcal{B}[L^2(X)]\subset\mathcal{L}[\mathcal{S}(X),\mathcal{S}'(X)].
\end{equation}
The conclusions of Section 50 in \cite{Treves} and the Corollary of Theorem 51.6 in \cite{Treves} imply that, isomorphically, 

\begin{equation}\label{Nucl-2}
\mathfrak{Int}:\mathcal{S}(X\times X)\overset{\sim}{\rightarrow}\mathcal{L}[\mathcal{S}'(X),\mathcal{S}(X)],\qquad 
\mathfrak{Int}:\mathcal{S}'(X\times X)\overset{\sim}{\rightarrow}\mathcal{L}[\mathcal{S}(X),\mathcal{S}'(X)].
\end{equation}

By putting together the informations above about the operations $K^A$ and $\mathfrak{Int}$, we get the following result concerning our 
functional calculus:

\begin{proposition}\label{jvartz}
If the potential vector $A$ is of class $C_{\text{pol}}^\infty$, the map $\mathfrak{Op}^A$ defines linear topological isomorphisms
$$
\mathfrak{Op}^A:\mathcal{S}(\Xi)\overset{\sim}{\rightarrow}\mathcal{L}[\mathcal{S}'(X),\mathcal{S}(X)],\qquad  
\mathfrak{Op}^A:\mathcal{S}'(\Xi)\overset{\sim}{\rightarrow}\mathcal{L}[\mathcal{S}(X),\mathcal{S}'(X)].
$$
\end{proposition}

So ``any'' operator is (in a unique way) a magnetic pseudodifferential operator of the form $\Op^A(f)$ for some tempered distribution $f$ and the 
regularizing operators are exactly those with symbol in the Schwartz space. 

Gauge covariance can be extended to this setting; we leave the details to the reader:

\begin{proposition}\label{gauge'}
Let $A$ and $A'$ two vector potentials of class $C_{\text{pol}}^\infty$ defining the same magnetic field, $dA=B=dA'$. Then there exists a real 
function $\rho\in C_{\text{pol}}^\infty(X)$ such that $A'=A+\nabla\rho$ and 
$e^{i\rho(Q)}\mathfrak{Op}^A(f)e^{-i\rho(Q)}=\mathfrak{Op}^{A+\nabla\rho}(f)\;$ for any $f\in\mathcal S'(\Xi)$; this second identity is valid in 
$\mathcal{L}[\mathcal{S}(X),\mathcal{S}'(X)]$.
\end{proposition}

\subsection{The magnetic Fourier-Wigner transformation and special classes of operators}

\begin{definition}
(a) For any pair of vectors $u,v$ from $\mathcal H=L^2(X)$ we define the function
\begin{equation}\label{mFW-def}
\mathcal{W}^A_{u,v}:\Xi\rightarrow\mathbb{C},\ \ \mathcal{W}^A_{u,v}(\xi):=<v,W^A(\xi)u>,
\end{equation}
called {\it the magnetic Fourier-Wigner transform of the couple} $(u,v)$. 

(b) The map $(v,u)\mapsto\mathcal{W}^A_{u,v}$ will be called {\it the magnetic Fourier-Wigner transformation} (defined by the vector potential 
$A$). 
\end{definition}

In fact $\Op^A(f)$ was defined by $\left<v,\Op^A(f)u\right>=\int_\Xi d\xi\;(\mathcal{F}_\Xi^{-1} f)(\xi)\mathcal{W}^A_{u,v}(\xi)$, $\ 
u,v\in\mathcal H$.

\begin{proposition}\label{propTFW}
(a) The magnetic Fourier-Wigner transformation extends to a unitary operator $\;\mathcal{W}^A:L^2(X\times X)\rightarrow L^2(\Xi)$.

(b) If $A$ is of class $C^\infty_{\text{pol}}$ then the magnetic Fourier-Wigner transformation defines isomorphisms $\mathcal{W}^A:\mathcal 
S(X\times X)\rightarrow \mathcal S(\Xi)$ and $\mathcal{W}^A:\mathcal S'(X\times X)\rightarrow \mathcal S'(\Xi)$.
\end{proposition}

\begin{proof}
Using the explicit form of $W^A(\xi)$ we obtain

\begin{equation}\label{mFW-form}
\mathcal{W}^A_{u,v}=\left[(\mathbf{1}\otimes\mathcal{F}_X)\mathfrak IS\left(\tilde{\Lambda}^A\right)^{-1}\right](u\otimes\overline{v}),
\end{equation}
where $\mathfrak I$ is the composition with the change of variables $(x,y)\mapsto(y,x)$ on $X\times X$. Under the right asumption, each of the 
maps $\mathbf{1}\otimes\mathcal{F}_X$, $\mathfrak I$, $S$ and $\left(\tilde{\Lambda}^A\right)^{-1}$ is an isomorphism between the 
corresponding spaces. One also uses the reinterpretation $\mathcal{W}^A_{u,v}\equiv\mathcal{W}^A_{u\otimes\overline{v}}$.
\end{proof}

An important direct consequence of this result is

\begin{corollary}
The Weyl system with magnetic field $\;W^A:\Xi\rightarrow\mathcal{U}[L^2(X)]$ is irreducible, i.e. there are no non-trivial subspaces of $L^2(X)$ 
invariant under all the operators $\{W^A(\xi)\mid\xi\in\Xi\}$.
\end{corollary}

\begin{proof}
Suppose that $\mathcal{K}$ is a closed non-trivial subspace of $L^2(X)$, invariant under all the operators $W^A(\xi)$, $\xi\in\Xi$. Let 
$v\in\mathcal{K}^\bot$ be different from $0$. Then for any $u\in\mathcal{K}\setminus\{0\}$ we have $W^A(\xi)u\in\mathcal{K}$ for any 
$\xi\in\Xi$, so that
$$
\mathcal{W}^A_{u,v}(\xi)=<v,W^A(\xi)u>=0,\qquad \forall \xi\in\Xi.
$$
Thus we deduce that $\|\mathcal{W}^A_{u,v}\|_{L^2(\Xi)}=0$. But $\|\mathcal{W}^A_{u,v}\|_{L^2(\Xi)}=\|u\|\|v\|$ and we get a contradiction.
\end{proof}

{\bf Remark.} The Fourier-Wigner transformation also serves to express the operators $\Op^A(F)$ in a convenient way. Let us stick, for 
example, to the case in which $A$ has tempered growth. Then for all $u,v\in\mathcal S(X)$ and $F\in\mathcal S'(\Xi)$, one has 
$\left<v,\Op^A(F)u\right>=\left<\overline{\mathcal{W}^A_{u,v}},\mathcal{F}_\Xi^{-1}F\right>$, the left-hand-side being interpreted as the 
anti-duality between $\mathcal S(X)$ and $\mathcal S'(X)$, while the right-hand-side as the anti-duality between $\mathcal S(\Xi)$ (cf. Prop. 
\ref{propTFW}, (b)) and $\mathcal S'(\Xi)$.

We shall identify now finite-rank, Hilbert-Schmidt and compact operators.

\begin{proposition}\label{clase}
(a) For any $u,v\in\mathcal H$ we have $|u><v|=\mathfrak{Op}^A\left(\mathcal F_\Xi\mathcal W^A_{u,v}\right)$.

(b) $\mathfrak{Op}^A$ induces a unitary map from $L^2(\Xi)$ to $\mathcal{B}_2(\mathcal{H})$, the ideal of Hilbert-Schmidt operators.

(c) The family $\Op ^A\left[\mathcal{F}_\Xi L^1(\Xi)\right]$ is dense in the closed ideal $\mathcal K(\H)$ of all compact operators in $\H$.
\end{proposition}

\begin{proof}
(a) The operator $|u><v|$ is an integral operator having the kernel $u\otimes\overline{v}$. Thus 
$$
|u><v|=\mathfrak{Int}(u\otimes\overline{v})=\mathfrak{Op}^A\left[(K^A)^{-1}(u\otimes \overline{v})\right].
$$
One has 
$$
(K^A)^{-1}(u\otimes \overline{v})=(\mathbf{1}\otimes\mathcal{F}_X)S(\tilde{\Lambda}^A)^{-1}(u\otimes 
\overline{v})=(\mathbf{1}\otimes\mathcal{F}_X)\mathfrak 
I(\mathbf{1}\otimes\overline{\mathcal{F}}_{X\star})(\mathbf{1}\otimes\mathcal{F}_X)\mathfrak IS(\tilde{\Lambda}^A)^{-1}(u\otimes 
\overline{v}). 
$$
But, by a simple calculation, one gets $(\mathbf{1}\otimes\mathcal{F}_X)\mathfrak 
I(\mathbf{1}\otimes\overline{\mathcal{F}}_{X\star})=\mathcal{F}_\Xi$. The point (a) follows by taking (\ref{mFW-form}) into account.

(b) On the intersection $\mathcal{F}_\Xi L^1(\Xi)\cap L^2(\Xi)$ we have $\mathfrak{Op}^A=\mathfrak{Int}\circ K^A$, where 
$K^A:L^2(\Xi)\rightarrow L^2(X\times X)$ is unitary  and $\mathfrak{Int}:L^2(X\times X)\rightarrow \mathcal{B}_2(\mathcal{H})$ is also unitary 
(a classical result). But $\mathcal{F}_\Xi L^1(\Xi)\cap L^2(\Xi)$ is dense in $L^2(\Xi)$.

(c) By (b), for all $f\in\mathcal{F}_\Xi L^1(\Xi)\cap L^2(\Xi)$ the operator $\Op^A(f)$ is Hilbert-Schmidt, hence compact. The space 
$\mathcal{F}_\Xi L^1(\Xi)\cap L^2(\Xi)$ is dense in $L^2(\Xi)$, thus $\Op^A\left[\mathcal{F}_\Xi L^1(\Xi)\cap L^2(\Xi)\right]$ is dense in $\mathcal 
B_2(\H)$ with respect to the Hilbert-Schmidt norm, hence also with respect to the operator norm. It follows that $\Op^A\left[\mathcal{F}_\Xi 
L^1(\Xi)\cap L^2(\Xi)\right]$ is dense in $\mathcal K(\H)$. But $\Op^A\left[\mathcal{F}_\Xi L^1(\Xi)\right]$ is also contained in $\mathcal K(\H)$, 
since $\mathcal{F}_\Xi L^1(\Xi)\cap L^2(\Xi)$ is dense in $\mathcal{F}_\Xi L^1(\Xi)$ and $\parallel\Op^A(f)\parallel_{\mathcal B(\H)}\le\parallel 
f\parallel_{\mathcal F_\Xi L^1}:=\parallel\mathcal F_\Xi^{-1}f\parallel_{L^1}$, $\forall f$.
\end{proof}

We see that $\Op^A$ has a strong tendency towards irreducibility: $\mathcal K(\H)$ is, of course, irreducible, thus, by density, 
$\Op^A\left[\mathcal{F}_\Xi L^1(\Xi)\cap L^2(\Xi)\right]$ is also an irreducible family of operators in $\H$. Other results of this type may be 
obtained by density.

\subsection{The correct form of the minimal coupling principle}

The loose form of the minimal coupling principle says that ``when a magnetic field $B=dA$ is turned on, one should replace the canonical variable 
$p$ with $p-A(x)$''. The question is, of course, at which stage should this replacement be performed when quantization of a classical observable 
$f$ is intended. The wrong answer is to compose $f:\Xi\rightarrow\mathbb C$ with the change of variables $(x,p)\mapsto(x,p-A(x))$ and then 
apply the Weyl calculus. As seen in the Introduction, this would give a gauge non-covariant formula. The right approach is to apply to $f$ itself a 
modified (magnetic) Weyl calculus. And this modification is governed actually by the sound, elementary form of the minimal coupling principle: the 
quantum observable $P$ is replaced by $\Pi^A=P-A(Q)$ and this object determines the expression of the Weyl system $W^A$, used in the 
definition of $f(Q,\Pi^A)$. One could say that this is correct, since $W^A$ summarizes the commutation relations of the family of operators 
$(Q_1,\dots,Q_N;\Pi^A_1,\dots,\Pi^A_N)$ for which a functional calculus is requested.

However, one could ask for a more sophisticated (and not so clear ideologically) form of the minimal coupling principle: find a transformation 
$T^A$ acting on phase-space functions such that, for any $f$, $f(Q,\Pi^A)$ is obtained (also) by Weyl quantizing the symbol $T^Af\equiv f^A$. A 
brief examination of this topic follows.

Let us assume, for convenience, that $B$ and $A$ are of class $C^\infty_{\text{pol}}$. Both $\Op^A$ and $\Op$ are one-to-one (even 
isomorphic) from $\mathcal S'(\Xi)$ to $\mathcal L[\mathcal S(X),\mathcal S'(X)]$. Using notations from Subsection 3.2, one has

$$
\Op^A(f)=\Op(f^A)\Leftrightarrow \mathfrak{Int}(K^Af)=\mathfrak{Int}(Kf^A)\Leftrightarrow f^A=K^{-1}K^Af.
$$
By using explicit formulae for $K^A$ and $K$ and the identity $S\tilde{\Lambda}^AS^{-1}=\tilde{\Lambda}^A\circ S$, one gets $f^A=T^Af$, with

$$
T^A:\mathcal S'(\Xi)\rightarrow\mathcal S'(\Xi),\ \ T^A:=\left(\mathbf 1\otimes\mathcal F_X\right)\left(\tilde{\Lambda}^A\circ S\right)\left(\mathbf 
1\otimes\overline{\mathcal F}_{X^\star}\right).
$$
Formally (or for suitable $f$'s)

$$
\left(T^Af\right)(x,p)=\int_X\int_{X^\star}dydk\;e^{iy\cdot\left[k-p+\int_{-1/2}^{1/2}dt
\;A(x+ty)\right]}f(x,k)=
$$
$$
=\int_Xdy\;e^{-iy\cdot\left[p-\int_{-1/
2}^{1/2}dt\;A(x+ty)\right]}(\mathbf 1\otimes\overline{\mathcal F}_{X^\star})f](x,y).
$$

One should compare this rather complicated formula (a sort of minimal coupling principle for all observables) with 
$$
\left(M^Af\right)(x,p):=f(x,p-A(x))=\int_X\int_{X^\star}dydk\;e^{iy\cdot\left[k-p+A(x)\right]}f(x,k).
$$
The rigorous expression behind this formal integral is

$$
M^A:\mathcal S'(\Xi)\rightarrow\mathcal S'(\Xi),\ \ M^A:=\left(\mathbf 1\otimes\mathcal F_X\right)\Sigma^A\left(\mathbf 
1\otimes\overline{\mathcal F}_{X^\star}\right),
$$
with $\Sigma^A(x,y)=e^{iy\cdot A(x)}$, $x,y\in X$.

A comparison of the explicit formulae for $T^A$ and $M^A$ shows once again the difference between the correct and the mistaken quantizations 
in a magnetic field: {\it The correct one involves circulations of the magnetic potential $A$ through segments $[x_1,x_2]$, while for the wrong one 
the same circulations are calculated by using the constant value} $A_{x_1,x_2}:=A\left(\frac{x_1+x_2}{2}\right)$, {\it taken at the middle of the 
respective segment}.

One has a complete characterization of the vector potentials for which the wrong quantization is good:

\begin{lemma}
One has $T^A=M^A$ (which is equivalent to $\Op^A(f)=\Op(M^Af)$, $\forall f\in\mathcal S'(\Xi)$) if and only if $A$ is linear.
\end{lemma}

\begin{proof}
$\mathbf 1\otimes\mathcal F_X$ and $\mathbf 1\otimes\overline{\mathcal F}_{X\star}$ being one-to-one, we have $T^A=M^A$ if and only if 
$\tilde{\Lambda}^A\circ S=\Sigma^A$, i.e. if and only if $\;y\cdot\int_{-1/2}^{1/2}dt\;\left[A(x+ty)-A(x)\right]$=0, $\forall x,y\in X$. A simple 
application of Taylor's formula shows that this is equivalent to the annulation of all the second derivatives of all the components of $A$.
\end{proof}

One of the most important examples is the constant magnetic field. In this case, everybody would choose a linear potential vector $A$ and no 
care is needed in the choice of the quantization procedure. Most articles involving a functional calculus in a magnetic field are written for 
constant $B$ {\it and} linear $A$. Note, however, that the identity $T^A=M^A$ {\it is not gauge invariant}.  

However, one can have $T^Af=M^Af$ for any $A$ for certain special functions $f$. This is obviously true if $f$ depends only on the variable $x\in 
X$. Actually, in this case $\Op^A(f)=\Op(M^Af)=f(Q)$. Let us give some more interesting examples. 

\begin{proposition}
Let $f$ be a polynomial of order $m$ in $p$, not depending on the variable in $X$. If $m\le2$, then $T^Af=M^Af$, hence $\Op^A(f)=\Op(M^Af)$. 
This is no longer true for $m=3$.
\end{proposition}

\begin{proof}
Let us consider the monomial $f_\alpha(x,p):=p^\alpha$. Then we have $(\mathbf 1\otimes\overline{\mathcal 
F}_{X^\star})f_\alpha=(-i\partial)^\alpha\delta$, thus $\left(T^Af\right)(x,p)=\left[(i\partial_y)^\alpha e^{-i\tau^A(x,p;y)}\right]|_{y=0}\;$ and 
$\left(M^Af\right)(x,p)=\left[(i\partial_y)^\alpha e^{-i\mu^A(x,p;y)}\right]|_{y=0}$, where the two phases are defined by 
$\tau^A(x,p;y):=y\cdot\left[p-\int_{-1/2}^{1/2}dt\;A(x+ty)\right]\ $ and $\mu^A(x,p;y):=y\cdot[p-A(x)]$. We concentrate on the cases $m=1,2,3$. 
The following list of relations is needed:

$$
i\partial_{y_j}e^{-i\varphi}=\left(\partial_{y_j}\varphi\right) e^{-i\varphi},\ \ \ 
i^2\partial_{y_k}\partial_{y_j}e^{-i\varphi}=\left(i\partial_{y_k}\partial_{y_j}\varphi+\partial_{y_k}\varphi\;\partial_{y_j}\varphi\right)e^{-i\varphi}, 
$$

$$
i^3\partial_{y_l}\partial_{y_k}\partial_{y_j}e^{-i\varphi}=
$$
$$
=\left(-\partial_{y_l}\partial_{y_k}\partial_{y_j}\varphi+i\partial_{y_j}\varphi\;\partial_{y_l}
\partial_{y_k}\varphi+i\partial_{y_k}\varphi\;\partial_{y_l}\partial_{y_j}\varphi+i\partial_{y_l}\varphi\;\partial_{y_k}\partial_{y_j}\varphi+\partial_{y_l}\
\varphi\;\partial_{y_k}\varphi\;\partial_{y_j}\varphi\right)e^{-i\varphi}.
$$

Note that $\partial_{y_j}\mu^A(x,p;y)=p_j-A_j(x)$, while the higher-order derivatives vanish.

A simple calculation gives 

$$
\partial_{y_j}\tau^A(x,p;y)=p_j-\int_{-1/2}^{1/2}dt\;A_j(x+ty)-\sum_{n=1}^Ny_n\int_{-1/2}^{1/2}tdt\;(\partial_jA_n)(x+ty),
$$
and by taking the value in $y=0$ one gets 
$\left[\partial_{y_j}\tau^A(x,p;y)\right]\vert_{y=0}=\left[\partial_{y_j}\mu^A(x,p;y)\right]\vert_{y=0}=p_j-A_j(x)$. Thus $T^Af=M^Af$ for any 
first-order polynomial. This is not amaizing: $\Op^A(p_j)=\Pi^A$ was accepted as a basic principle.

One also has 

$$
\partial_{y_k}\partial_{y_j}\tau^A(x,p;y)=-\int_{-1/2}^{1/2}tdt\;(\partial_jA_k+\partial_kA_j)(x+ty)-\sum_{n=1}^Ny_n\int_{-1/2}^{1/2}t^2dt\;
(\partial_k\partial_jA_n)(x+ty).
$$
By ``miracle'' this term vanishes in $y=0$; then straightforwardly $T^Af=M^Af$ also for second-order polynomials. This is significant, since most 
of the time people considered the case $f(x,p)=\vert p\vert^2$, leading to the magnetic Laplacian $\Delta^A=(\Pi^A)^2$; no care is needed in this 
case.

The situation changes drastically for third order polynomials. One has

$$
\partial_{y_l}\partial_{y_k}\partial_{y_j}\tau^A(x,p;y)=-\int_{-1/2}^{1/2}t^2dt\;\left(\partial_k\partial_jA_l+\partial_l\partial_jA_k+\partial_l\partial_k
A_j\right)(x+ty)-
$$
$$
-\sum_{n=1}^Ny_n\int_{-1/2}^{1/2}t^3dt\;(\partial_l\partial_k\partial_jA_n)(x+ty),
$$
which in $y=0$ takes the value $-\frac{1}{12}\left(\partial_k\partial_jA_l+\partial_l\partial_jA_k+\partial_l\partial_kA_j\right)(x)$. In this case 
$T^Af\ne M^Af$.

\end{proof}

\section{The magnetic Moyal algebra}

We come now to an important point in the development of our functional calculus. The product of the operators $\mathfrak{Op}^A(f)$ and 
$\mathfrak{Op}^A(g)$ is again an integral operator with a kernel $K^A(f\circ^B g)$, that can formally be associated to the function on $\Xi$ 
obtained by the following non-commutative composition law, called {\it the magnetic Moyal product of the functions} $f$ {\it and} $g$:

\begin{equation}\label{circ-def}
(f\circ^B g)(\xi)=4^N\int_\Xi d\eta \int_\Xi d\zeta \; e^{-2i\sigma(\xi-\eta,\xi-\zeta)}e^{-i\Gamma^B(<q-y+x,x-q+y,y-x+q>)}f(\eta)g(\zeta)=
\end{equation}
$$
=4^N\int_\Xi d\eta \int_\Xi d\zeta \; e^{-2i\sigma(\eta,\zeta)}e^{-i\Gamma^B(<q-x-y,q+x-y,q+y-x>)}f(\xi-\eta)g(\xi-\zeta);
$$
here $\xi=(q,p),\;\eta=(x,k),\; \zeta=(y,l)$. Thus we have $\mathfrak{Op}^A(f\circ^B g)=\mathfrak{Op}^A(f)\mathfrak{Op}^A(g)$. We can also 
define an involution (the same as in the non-magnetic case) by $f^\circ(\xi):=\overline{f(\xi)}$ such that 
$\mathfrak{Op}^A(f^\circ)=\mathfrak{Op}^A(f)^*$.

The integral defining $f\circ^B g$ is absolutely convergent only for restricted classes of symbols. One seeks to extend the composition law 
$\circ^B$ to large classes of distributions in such a way as to obtain (together with the involution $^\circ$) $^*$-algebras.  For any choice of a 
magnetic potential, the functional calculus with magnetic field will be a representation of these $^*$-algebras. But the algebras themselves are 
completely intrinsic, being defined only in terms of the magnetic field. We shall do this extension by duality, following the approach of 
\cite{Antonets} and \cite{Gracia1} (see also \cite{Gracia2} and \cite{Estrada}) valid in the absence of the magnetic field. We obtain a magnetic 
analog of the Moyal algebra outlined in Subsection 1.4; the terminology is that of the references above and it is suggested by some early 
fundamental work of Moyal (cf. \cite{Moyal}). 

The standard technique of extending the composition law, based on oscillatory integrals and classes of symbols, is less appropriate for our 
present purposes. But we intend to deal with this topic in a subsequent publication.

\subsection{The magnetic Moyal product}
 
Before discussing rigorously the sense of formula (\ref{circ-def}) for various assumptions on $f$, $g$ and $B$, we make some formal remarks. 
Note that if $B=0$, (\ref{circ-def}) reduces to the usual composition of symbols (\ref{Moya}) in the Weyl quantization. The magnetic correction 
consists of a phase factor defined in terms of the flux of the magnetic field through suitable triangles. The associativity of the above composition 
law comes from the 2-cocycle condition, the second identity in (\ref{trebuie}). For this just notice that 
$e^{-i\Gamma^B(<q-y+x,x-q+y,y-x+q>)}=\Omega^B(q-y+x;2y-2q,2q-2x)$ and do the right calculation. Finally, it is easy to check that 
$\mathfrak{Op}^A(f\circ^B g)=\mathfrak{Op}^A(f)\mathfrak{Op}^A(g)$ whenever everything is well-defined.

In general, if no special assumption on $B$ is imposed, it is not so easy to define and use sharply the magnetic Moyal product. For $f,g\in 
\mathcal F_\Xi L^1(\Xi)$ both $\mathfrak{Op}^A(f)$ and $\mathfrak{Op}^A(g)$ are defined as bounded linear operators in $\mathcal H=L^2(X)$, 
but it is not clear if their product is of the form $\mathfrak{Op}^A(h)$ for some $h$ (eventually in $\mathcal F_\Xi L^1(\Xi)$). On the other hand, 
if $f,g\in L^1(\Xi)$, then the integral in (\ref{circ-def}) is absolutely convergent and defines a bounded continuous function on $\Xi$. However, we 
do not see why should this function be integrable and, anyway, applying $\Op^A$ to all these is problematic. One can also take advantage of 
Proposition \ref{clase} (b) to endow $L^2(\Xi)$ with the structure of a $^*$-algebra, the composition law coinciding with (\ref{circ-def}) on 
suitable subsets.

In fact many other solutions exist but {\it they do not seem to be natural enough in the present framework}. In a future publication we will give 
another (equivalent) form of the magnetic Moyal product (a ``very twisted convolution'') for which it will be relatively easy to define nice 
normed $^*$-algebras. Their pull-backs in the pseudo-differential representation involve a partial Fourier transformation which has not an 
explicitly expressible range. So we postpone the study of magnetic composition laws for general (continuous) magnetic fields and turn to a 
special case which is, however, very comprehensive.

\begin{proposition}\label{bine}
Assume that the components of the magnetic field $B$ are of class $C^\infty_{\text{pol}}$. 

(a) For any $f,g\in\mathcal S(\Xi)$ one has $f\circ^B g\in\mathcal S(\Xi)$. The map $\circ^B:\mathcal S(\Xi)\times\mathcal 
S(\Xi)\rightarrow\mathcal S(\Xi)$ is bilinear and continuous.

(b) For any continuous vector potential $A$ such that $dA=B$, one has $\;\mathfrak{Op}^A(f\circ^B g)=\mathfrak{Op}^A(f)\mathfrak{Op}^A(g)$.
\end{proposition}

\begin{proof}
One can prove that $\mathcal S(\Xi)\circ^B\mathcal S(\Xi)\subset\mathcal S(\Xi)$ directly, estimating $\xi^\alpha\partial^\beta(f\circ^Bg)$ by 
involved manipulations in (\ref{circ-def}). This also gives the required continuity. We prefer to outline a simpler proof, based on integral kernels.

Let us choose $A\in C^\infty_{\text{pol}}(X,X^\star)$ such that $dA=B$; this is possible by (\ref{transversal}). We know that the maps 
$K^A:=\tilde{\Lambda}^AS^{-1}(\mathbf{1}\otimes\overline{\mathcal{F}}_{X^\star}):\mathcal S(\Xi)\rightarrow\mathcal S(X\times X^\star)$, $\ 
\mathfrak{Int}:\mathcal S(X\times X^\star)\rightarrow\mathcal L[\mathcal S'(X),\mathcal S(X)]$ and $\Op^A:\mathcal S(\Xi)\rightarrow\mathcal 
L[\mathcal S'(X),\mathcal S(X)]$ are topological isomorphisms and that $\Op^A=\mathfrak{Int}\circ K^A$.  One checks easily that $K^A(f\circ^B 
g)=K^A(f)\ddagger K^A(g)$, where $(L\ddagger M)(x,y):=\int_Xdz\;L(x,z)M(z,y)$ is the composition rule of integral kernels (leading to the 
product of the integral operators involved). These facts and the continuity of $\ddagger:\mathcal S(X\times X^\star)\times\mathcal S(X\times 
X^\star)\rightarrow\mathcal S(X\times X^\star)$ imply both the points (a) and (b) for $A\in C^\infty_{\text{pol}}(X,X^\star)$. 

The general case of a continuous $A$ in (b) is solved by gauge covariance, cf. Proposition \ref{gaco} (we consider here $\mathcal L[\mathcal 
S'(X),\mathcal S(X)]$ embedded in $\mathcal B\left[L^2(X)\right]$).
\end{proof}

Since $\mathcal S(\Xi)$ is obviously stable under involution, $(\mathcal S(\Xi),\circ^B,^\circ)$ is a $^*$-algebra (all the axioms are easily 
verified) and $\Op^A:\mathcal S(\Xi)\rightarrow\mathcal L[\mathcal S'(X),\mathcal S(X)]$ is a $^*$-isomorphism. Unfortunately, $\mathcal 
S(\Xi)$ is too small for many purposes. For instance, functions depending only on $x$ or on $p$ are not included.

\subsection{Extension by duality}

We start extending by duality the magnetic Moyal product with an asymmetric version: we shall compose a Schwartz test function with a 
tempered distribution. The result is a priori a tempered distribution, but we shall be able to get more precise informations in certain cases. The 
components of the magnetic field will be always considered to be in $C^\infty_{\text{pol}}$, thus the conclusions of Theorem \ref{bine} hold.

The duality approach is facilitated by the next Lemma:

\begin{lemma}\label{trace}
For any functions $f$ and $g$ in $\mathcal{S}(\Xi)$ we have
$$
\int_\Xi d\xi\;(f\circ^B g)(\xi)=\int_\Xi d\xi\;(g\circ^B f)(\xi)=\int_\Xi d\xi\;f(\xi)g(\xi)=<\overline{f},g>\equiv(f,g).
$$
\end{lemma}

\begin{proof}
Of course, one needs only to show that $\int_\Xi d\xi\;(f\circ^B g)(\xi)=\int_\Xi d\xi\;f(\xi)g(\xi)$; the other identities are trivial consequences.

The calculation will be straightforward by regularization. We choose sequences $(a_n)_{n\in\mathbb N}\in\mathcal S(X)$, $(b_n)_{n\in\mathbb 
N}\in\mathcal S(X^\star)$ such that $a_n\rightarrow 1$ in $\mathcal S'(X)$ and $b_n\rightarrow 1$ in $\mathcal S'(X^\star)$. Then it is shown 
easily that 

$$
\int_X\int_{X^\star}dq\;dp\;(f\circ^B g)(q,p)a_n(q)b_m(p)\rightarrow\int_X\int_{X^\star}dq\;dp\;f(q,p)g(q,p)
$$
for $n,m\rightarrow\infty$. For this we use the explicit formula for $f\circ^B g$, Fubini's Theorem, the fact that the Fourier transforms of $a_n$ 
and $b_m$ converge respectively to the distribution $\delta$ and the annulation of $\Gamma^B(<q-y+x,x-q+y,y-x+q>)$ for $x=y$. We leave the 
details to the reader.
\end{proof}

\begin{corollary}\label{circ-dual}
For any three functions $f$, $g$ and $h$ in $\mathcal{S}(\Xi)$ we have
$$
(f\circ^B g,h)=(f,g\circ^B h)=(g,h\circ^B f).
$$
\end{corollary}

\begin{proof}
Easy consequence of the Lemma, the associativity of $\circ^B$ and the symmetry of $(\cdot,\cdot)$.
\end{proof}

\begin{definition}\label{circ-ext}
For any distribution $F\in\mathcal{S}'(\Xi)$ and any function $f\in\mathcal{S}(\Xi)$ we define
$$
(F\circ^B f,h):=(F,f\circ^B h),\qquad (f\circ^B F,h):=(F,h\circ^B f),\qquad \forall h\in\mathcal{S}(\Xi).
$$
\end{definition}

By using Proposition \ref{bine} (a) and the Definition it is straightforward to see that

\begin{proposition}
The above definition provides two bilinear continuous mappings $\mathcal{S}'(\Xi)\times\mathcal{S}(\Xi)\rightarrow\mathcal{S}'(\Xi)$, resp. 
$\mathcal{S}(\Xi)\times\mathcal{S}'(\Xi)\rightarrow\mathcal{S}'(\Xi)$.
\end{proposition}

One easily checks that $(F\circ^B g)^\circ=g^\circ\circ^B F^\circ$ and $(g\circ^B F)^\circ=F^\circ\circ^B g^\circ$, for all $F\in\mathcal S'(\Xi)$ 
and $g\in\mathcal S(\Xi)$. Associativity results as $(f_1\circ^B F)\circ^B f_2=f_1\circ^B (F\circ^B f_2)$, for $f_1,f_2\in\mathcal S(\Xi)$, 
$F\in\mathcal S'(\Xi)$ obviously hold, so one can define unambiguously $f_1\circ^B\cdots\circ^Bf_n$ if one $f_j$ is a tempered distribution and all 
the others are Schwartz test functions. Lemma \ref{trace} implies immediately that $1\circ^B f=f=f\circ^B 1$, $\ \forall f\in\mathcal S(\Xi)$.

\begin{proposition}\label{chestie}
For any vector potential $A$ with tempered growth, $\Op^A$ is an involutive linear continuous map $:\mathcal{S}'(\Xi)\mapsto\mathcal 
L[\mathcal S(X),\mathcal S'(X)]$, satisfying  $\Op^A(F\circ^B g)=\Op^A(F)\Op^A(g)$ and $\Op^A(g\circ^B F)=\Op^A(g)\Op^A(F)$ for all 
$F\in\mathcal S'(\Xi)$ and $g\in \mathcal S(\Xi)$.
\end{proposition}

\begin{proof}
We already know that $\Op^A:\mathcal{S}'(\Xi)\mapsto\mathcal L[\mathcal S(X),\mathcal S'(X)]$ is an isomorphism of topological vector 
spaces. The involution on $\mathcal L[\mathcal S(X),\mathcal S'(X)]$ is defined by antiduality ($\left<T^*v,u\right>=\left<v,Tu\right>$, $\forall 
u,v\in\mathcal S(X)$). Then the formula $\Op^A(F)^*=\Op^A(F^\circ)$ follows readily from the Remark in Subsection 3.3. The relations 
$\Op^A(F\circ^B g)=\Op^A(F)\Op^A(g)$ and $\Op^A(g\circ^B F)=\Op^A(g)\Op^A(F)$ follow by approximating $F\in\mathcal S'(\Xi)$ with 
elements $f_n$ of $\mathcal S(\Xi)$; all the continuity properties which are needed are already proved.

\end{proof}

\subsection{The magnetic Moyal $^*$-algebras}

\begin{definition}\label{Moyal} (a) The spaces of distributions

\begin{equation}\nonumber
\mathcal{M}_L(\Xi):=\left\{F\in\mathcal{S}'(\Xi)\;\mid\;F\circ^B f\in\mathcal{S}(\Xi),\;\;\forall f\in\mathcal{S}(\Xi)\right\}
\end{equation}
and
\begin{equation}\nonumber
\mathcal{M}_R(\Xi):=\left\{F\in\mathcal{S}'(\Xi)\;\mid\;f\circ^B F\in\mathcal{S}(\Xi),\;\;\forall f\in\mathcal{S}(\Xi)\right\}
\end{equation}
will be called, respectively, {\it the left} and {\it the right magnetic Moyal algebra}.

(b) Their intersection

$$
\mathcal{M}(\Xi):=\mathcal{M}_L(\Xi)\cap\mathcal{M}_R(\Xi)
$$
will be called {\it the magnetic Moyal algebra}.
\end{definition}

The three spaces above depend on the magnetic field so, in principle, they would deserve an index $B$. 

For any two distributions $F$ and $G$ in $\mathcal{M}(\Xi)$ we can extend the magnetic Moyal product by

$$
(F\circ^B G,h):=(F,G\circ^B h),\qquad \forall h\in\mathcal{S}(\Xi).
$$

\begin{proposition}\label{structMoy}
The set $\mathcal{M}(\Xi)$ together with the composition law $\circ^B$ defined as above and the complex conjugation $F\mapsto F^\circ$ is an 
unital $^*$-algebra, containing $\mathcal S(\Xi)$ as a self-adjoint two-sided ideal.
\end{proposition}

All the verifications are trivial. Since the constant functions are obviously in $\mathcal{M}(\Xi)$, it is already clear that the $^*$-algebra 
$\mathcal S(\Xi)$ is enlarged. We shall see in Subsection 4.4 that this enlargement is substantial.

We study now the behaviour of $\Op^A$ on symbols belonging to the magnetic Moyal algebra.

\begin{proposition}\label{imopMoy}
$\Op^A$ is an isomorphism of $\ ^*$-algebras betweeen $\mathcal{M}(\Xi)$ and $\mathcal L[\mathcal S(X)]\cap\mathcal L[\mathcal S'(X)]$.
\end{proposition}

\begin{proof}
Let us denote simply $\mathcal S=\mathcal S(X)$ and $\mathcal S'=\mathcal S'(X)$. We identify $\mathcal L(\mathcal S)\equiv\mathcal 
L(\mathcal S,\mathcal S)$ with the family of all the elements $T\in\mathcal L(\mathcal S,\mathcal S')$ such that $T\mathcal S\subset\mathcal 
S$. By the Closed Graph Theorem, such a $T$ will automatically be continuous (and linear) as a mapping $\mathcal S\mapsto\mathcal S$. 
$\;\mathcal L(\mathcal S)$ is obviously an algebra with the composition of operators.

Another algebra is $\mathcal L(\mathcal S')\equiv\mathcal L(\mathcal S',\mathcal S')$, which may be identified to the family of elements 
$T\in\mathcal L(\mathcal S,\mathcal S')$ that admit a continuous extension to $\mathcal S'$. We recall that the involution on $\mathcal 
L(\mathcal S,\mathcal S')$ is defined by antiduality ($\left<T^*u,w\right>=\left<u,Tw\right>$, $\forall u,w\in\mathcal S$). Then, plainly, $\mathcal 
L(\mathcal S)^*=\mathcal L(\mathcal S')$ and $\mathcal L(\mathcal S')^*=\mathcal L(\mathcal S)$. Thus $\mathcal L(\mathcal S)\cap\mathcal 
L(\mathcal S')$ is a $^*$-algebra.

We know that $\Op^A$ is one-to-one; we calculate now $\Op^A[\mathcal{M}(\Xi)]$. By the definition of $\mathcal{M}(\Xi)$ and Proposition 
\ref{jvartz}, $T$ is in $\Op^A[\mathcal{M}(\Xi)]$ if and only if $T\mathcal L(\mathcal S',\mathcal S)\subset\mathcal L(\mathcal S',\mathcal S)$ 
and $\mathcal L(\mathcal S',\mathcal S)T\subset\mathcal L(\mathcal S',\mathcal S)$. The last inclusion is equivalent to $T^*\mathcal L(\mathcal 
S',\mathcal S)\subset\mathcal L(\mathcal S',\mathcal S)$. It is easy to see that $T\mathcal L(\mathcal S',\mathcal S)\subset\mathcal 
L(\mathcal S',\mathcal S)$ if and only if $T\in\mathcal L(\mathcal S)$. One implication is trivial and the other one follows once again from the 
Closed Graph Theorem and from the fact that for any $v\in\mathcal S$ there exist $u\in S'$ and $S\in\mathcal L(\mathcal S',\mathcal S)$ such 
that $Su=v$. Then, by taking also $T^*$ into account, we see that $\Op^A[\mathcal{M}(\Xi)]=\mathcal L(\mathcal S)\cap\mathcal L(\mathcal 
S')$.

Let $F,G\in\mathcal{M}(\Xi)$. We calculate for $u\in\mathcal S'$ and $h\in\mathcal S(\Xi)$
$$
\Op^A(F\circ^B G)\left[\Op^A(h)u\right]=\Op^A(F\circ^B G\circ^B h)u=
$$
$$
=\Op^A(F)\left[\Op^A(G\circ^B 
h)u\right]=\left[\Op^A(F)\Op^A(G)\right]\Op^A(h)u,
$$
where we used Proposition \ref{chestie}. Since any $v\in\mathcal S$ can be written as $\Op^A(h)u$ for some $u\in\mathcal S'$ and 
$h\in\mathcal S(\Xi)$, the multiplicative property of $\Op^A$ on $\mathcal{M}(\Xi)$ is shown.

The involutivity of $\Op^A$ is valid on $\mathcal S'(\Xi)$, as remarked before.
\end{proof}

{\bf Remark.} Propositions \ref{structMoy} and \ref{imopMoy} are the most important results. We note here rapidly some extra results concerning 
the magnetic Moyal algebras, all of an elementary nature. One also defines by duality products of the form $F_1\circ^B G\circ^B F_2\in\mathcal 
S'(\Xi)$ for $F_1\in\mathcal{M}_R(\Xi)$, $F_2\in\mathcal{M}_L(\Xi)$ and $G\in\mathcal S'(\Xi)$; $\mathcal S'(\Xi)$ is a 
$(\mathcal{M}_R(\Xi),\mathcal{M}_L(\Xi))$-bimodule. In fact $\mathcal{M}_L\circ^B\mathcal{M}_L\subset\mathcal{M}_L$ and 
$\mathcal{M}_R\circ^B\mathcal{M}_R\subset\mathcal{M}_R$, hence $\mathcal{M}_L$ and $\mathcal{M}_R$ are algebras. But they are different 
and correspond to each other by complex conjugation, so $\mathcal{M}$ is optimally defined as a $^*$-algebra by the present methods. The 
proof of Proposition \ref{imopMoy} also leads to $\Op^A\mathcal{M}_L(\Xi)=\mathcal L(\mathcal S)$ and $\Op^A\mathcal{M}_R(\Xi)=\mathcal 
L(\mathcal S')$.

The next striking result shows once more the importance of the magnetic Moyal algebras.

\begin{proposition}\label{egalitati}
One has $\mathcal S'(\Xi)\circ^B\mathcal S(\Xi)\subset\mathcal{M}_R(\Xi)$ and $\mathcal S(\Xi)\circ^B\mathcal 
S'(\Xi)\subset\mathcal{M}_L(\Xi)$.
\end{proposition}
\begin{proof}
$$
\Op^A[\mathcal S'(\Xi)\circ^B\mathcal S(\Xi)]=\Op^A[\mathcal S'(\Xi)]\Op^A[\mathcal S(\Xi)]=\mathcal L(\mathcal S,\mathcal S')\mathcal 
L(\mathcal S',\mathcal S)=\mathcal L(\mathcal S')\subset\Op^A\mathcal{M}_R(\Xi),
$$
thus $\mathcal S'(\Xi)\circ^B\mathcal S(\Xi)\subset\mathcal{M}_R(\Xi)$. The other inclusion is proved analogously.
\end{proof}

We note that both the inclusions are strict. For zero magnetic field $f\circ G$ is smooth if $f\in\mathcal S(\Xi)$ and $G\in\mathcal S'(\Xi)$, cf. 
\cite{Gracia1}.

\subsection{Some important subclasses}

We keep the setting of the preceding paragraphs, i.e. the components of $B$ (and those of $A$ when necessary) are of class 
$C_{\text{pol}}^\infty$. Simple examples show readily that $\mathcal{M}(\Xi)$ is much larger than $\mathcal S(\Xi)$. One shows easily that if 
$f(x,p)=f_1(x)$ depends only on the variable in $X$, then $\Op^A(f)=f_1(Q)$. If $f_1$ has tempered growth then $f_1(Q)\in\mathcal L(\mathcal 
S)\cap\mathcal L(\mathcal S')$, thus $f\in\mathcal{M}(\Xi)$ by Proposition \ref{imopMoy}. It is also quite obvious that $\mathcal F_\Xi 
L^1(\Xi)\subset\mathcal{M}(\Xi)$, since $W^A(\xi)$ is a continuous operator in $\mathcal S$ for all $\xi\in\Xi$. Actually, the same argument would 
also show that Fourier transforms of bounded, complex measures on $\Xi$ are also in the magnetic Moyal algebra. In the sequel we shall outline a 
less evident example. 

Let $C^\infty_{\text{\rm pol,u}}(\Xi)\subset\mathcal{S}'(\Xi)$ be the space of indefinitely derivable complex functions on $\Xi$ having uniform 
polynomial growth at infinity; i.e. $f\in C^\infty_{\text{\rm pol,u}}(\Xi)$ when it is indefinitely derivable and there exists $m\in\mathbb{N}$ 
(depending on $f$) such that for any multi-index $a\in\mathbb{N}^{2N}$ one has $|(\partial^a f)(\xi)|\leq C_a<\xi>^{m}$ for all $\xi\in\Xi$.

\begin{proposition}
$C^\infty_{\text{\rm pol,u}}(\Xi)\subset\mathcal{M}(\Xi)$.
\end{proposition}

\begin{proof}
First we reduce our proof to a precise estimate. Classes of functions $\mathcal T(\Xi)$ will be denoted briefly by $\mathcal T$.

For any $m\in\mathbb R$ we set $S^m_0:=\{f\in C^\infty\mid \;<\cdot>^{-m}\partial^a f\in L^\infty,\;\forall a\in\mathbb N^{2N}\}$ and 
$R^m_0:=\{f\in C^\infty\mid <\cdot>^{-m}\partial^a f\in L^1,\;\forall a\in\mathbb N^{2N}\}$. The reason for introducing these function spaces is the 
fact that $C^\infty_{\text{\rm pol,u}}(\Xi)=\cup_{m\in\mathbb R}S^m_0=\cup_{m\in\mathbb R}R^m_0$. This follows from $R^m_0\subset 
S^m_0\subset R^{m+2N+\varepsilon}_0$, valid for any $m\in\mathbb R$ and any $\varepsilon>0$, which is shown by trivial estimates. So the 
Proposition will be proved if we show that $R^m_0\subset\mathcal M$, $\forall m\in\mathbb R$. 

In fact the spaces $S^m_0$ were introduced only for comparison. They are the first constituents of $C^\infty_{\text{\rm pol,u}}$ you would think 
of, but technically the classes $R^m_0$ are better suited, since $L^1$-inequalities in initial spaces are within reach (for example we have 
$\parallel f\circ^Bg\parallel_\infty\;\le\parallel f\parallel_1\parallel g\parallel_1$). $R^m_0$ is a locally convex space with the family of norms 
$\{r^m_n\}_{n\in N}$, where $r^m_n(f):=\sum_{\vert a\vert\le n}\parallel\partial^a[<\cdot>^{-m}f]\parallel_1$ (other, equivalent, family is obtained 
by writing $\partial^a$ and $<\cdot>^{-m}$ in reversed order.)

Since $\mathcal S$ is dense in $R^m_0$, to show that $R^m_0\subset\mathcal M$ it will be sufficient to prove that for any $g\in \mathcal S$ the 
mappings $\mathcal S\ni f\mapsto f\circ^Bg\in\mathcal S$ and $\mathcal S\ni f\mapsto g\circ^Bf\in\mathcal S$ are continuous if on the initial 
space $\mathcal S$ we cosider the topology induced from $R^m_0$. We shall treat the first mapping; in fact this is enough, since $R^m_0$ is left 
invariant by complex conjugation. 

In the sequel we shall always write $\xi=(q,p)$, $\eta=(x,k)$ and $\zeta=(y,l)$. We also set 
$\phi^B(q,x,y):=e^{-i\Gamma^B(<q-y+x,x-q+y,y-x+q>)}$; it is a function in $C_{\text{pol}}^\infty$ with $\vert\phi^B(q,x,y)\vert=1$. By taking 
into account the discussion above and the form of the seminorms on $\mathcal S$, we see that it is enough to prove that for any $g\in\mathcal 
S$, $\alpha,\beta,\gamma,\delta\in\mathbb N^N$ and $m\in\mathbb R$ (I shall take $m$ to be an even positive integer) there exist $n\in\mathbb 
N$ and $C<\infty$ (they both depend on everything) such that 

$$
\parallel q^\alpha p^\beta\partial_q^\gamma\partial_p^\delta(f\circ^B g)\parallel_\infty\;\le Cr^m_n(f).
$$ 
Now the proof will proceed in several steps:

{\it Step 1.} A simple calculation gives 
$$
\left[q^\alpha p^\beta\partial_q^\gamma\partial_p^\delta(f\circ^B g)\right](q,p)=
$$
$$
=\sum_{\gamma'\le\gamma}C^{\alpha\beta\gamma\delta}_{\gamma'}\int_{\Xi}\int_{\Xi}d\eta\;d\zeta\;q^\alpha p^\beta(k-l)^{\gamma'}(y-x)^\delta 
e^{-2i\sigma(\xi-\eta,\xi-\zeta)}(\partial_q^{\gamma-\gamma'}\phi^B)(q,x,y)f(\eta)g(\zeta).
$$
The Leibnitz rule was used, as well as the two identities 
$$
\partial_p^\delta e^{-2i\sigma(\xi-\eta,\xi-\zeta)}=(2i)^{\vert\delta\vert}(y-x)^\delta 
e^{-2i\sigma(\xi-\eta,\xi-\zeta)}
$$ 
$$
\partial_q^{\gamma'}e^{-2i\sigma(\xi-\eta,\xi-\zeta)}=(2i)^{\vert\gamma'\vert}(k-l)^{\gamma'}e^{-2i\sigma(\xi-\eta,\xi-\zeta)}.
$$
{\it Step 2.} Having in view the form of $r^m_n$, we write $f(\eta)=<\eta>^m\left[<\eta>^{-m}f(\eta)\right]\;$ ($m$ even). By developping, the 
factor $<\eta>^m$ contributes with terms of the form $x^\mu k^\nu$. Thus, we need to estimate objects as 
$$
\int_{\Xi}\int_{\Xi}d\eta\;d\zeta\;q^{\alpha_1}x^{\alpha_2}(y-x)^{\alpha_3}p^{\alpha_4}k^{\alpha_5}(k-l)^{\alpha_6}e^{-2i\sigma(\xi-\eta,\xi-\zeta)}
\varphi(q,x,y)\left[<\eta>^{-m}f(\eta)\right]g(\eta).
$$ 
Here $\varphi$ is $C_{\text{pol}}^\infty$ in all the variables; the zero order derivative is no longer bounded and this will cause some 
complications.

{\it Step 3.} The heart of the proof lies in exploiting the nice properties of the factor $e^{-2i\sigma(\xi-\eta,\xi-\zeta)}$ by integrations by parts 
(an oscillatory integral technique). This works efficiently only with respect to certain of the variables. We produce these variables by making 
linear combinations of other, non-convenient ones. Let us write, for instance, $p=(p-l)+l$, $k-l=(k-p)+(p-l)$ and $k=(k-l)+l=(k-p)+(p-l)+l$. 
Then, plainly, we are reduced to estimating terms of the form

$$
\int_{\Xi}\int_{\Xi}d\eta\;d\zeta\;q^{\beta_1}x^{\beta_2}(y-x)^{\beta_3}l^{\beta_4}(p-k)^{\beta_5}(p-l)^{\beta_6}e^{-2i\sigma(\xi-\eta,\xi-\zeta)}\
varphi(q,x,y)\left[<\eta>^{-m}f(\eta)\right]g(\eta).
$$

{\it Step 4.} We use 
$$
(p-k)^{\beta_5}(p-l)^{\beta_6}e^{-2i\sigma(\xi-\eta,\xi-\zeta)}=(2i)^{-\vert\beta_5\vert}(-2i)^{-\vert\beta_6\vert}\partial_y^{\beta_5}\partial_x
^{\beta_6}\left[e^{-2i\sigma(\xi-\eta,\xi-\zeta)}\right].
$$ 
After an integration by parts and an application of Leibnitz's rule we see that we are reduced to bound terms as 
$$
\int_{\Xi}\int_{\Xi}d\eta\;d\zeta\;\psi(q,x,y)e^{-2i\sigma(\xi-\eta,\xi-\zeta)}\partial_x^{\gamma_1}\left[<\eta>^{-m}f(\eta)\right]l^{\gamma_2}\left(\
partial_y^{\gamma_3}g\right)(\zeta)
$$
for some unbounded $C_{\text{pol}}^\infty$-function $\psi$. Actually, only the polynomial estimate on $\psi$ itself will count now.

{\it Step 5.} Polynomial bounds are very democratic with respect to the choice of variables; use for example inequalities of the form 
$<x+y>\le2^{1/2}<x><y>$. Thus we can write 
$$
\psi(q,x,y)=\left[\frac{\psi(q,x,y)}{<y>^j<q-x>^j<q-y>^j}\right]\;\left[<y>^j<q-x>^j<q-y>^j\right]
$$ 
and the first factor will be bounded for $j$ large enough. By developping, we need to estimate 
$$
\int_{\Xi}\int_{\Xi}d\eta\;d\zeta\;\rho(q,x,y)(q-x)^{\delta_1}(q-y)^{\delta_2}e^{-2i\sigma(\xi-\eta,\xi-\zeta)}\partial_x^{\delta_3}\left[<\eta>^{-m}f
(\eta)\right]y^{\delta_4}l^{\delta_5}\left(\partial_y^{\delta_6}g\right)(\zeta),
$$
where now $\rho$ is bounded.

{\it Step 6.} But one has 
$$
(q-x)^{\delta_1}(q-y)^{\delta_2}e^{-2i\sigma(\xi-\eta,\xi-\zeta)}=(-2i)^{-\vert\delta_1\vert}(2i)^{-\vert\delta_2\vert}\partial_l^{\delta_1}\partial
_k^{\delta_2}e^{-2i\sigma(\xi-\eta,\xi-\zeta)}.
$$
We perform our last integration by parts, reducing ourselves to estimate 
$$
\int_{\Xi}\int_{\Xi}d\eta\;d\zeta\;\rho(q,x,y)e^{-2i\sigma(\xi-\eta,\xi-\zeta)}\partial_k^{\delta_2}\partial_x^{\delta_3}\left[<\eta>^{-m}f(\eta)\right]y
^{\delta_4}\left[\partial_l^{\delta_1}\partial_y^{\delta_6}\left(l^{\delta_5}g\right)\right](\zeta).
$$

{\it Step 7.} Obviously, this integral is dominated for any $\xi$ by 
$\parallel\partial_k^{\delta_2}\partial_x^{\delta_3}\left[<\cdot>^{-m}f\right]\parallel_1\;\parallel 
y^{\delta_4}\left[\partial_l^{\delta_1}\partial_y^{\delta_6}\left(l^{\delta_5}g\right)\right]\parallel_1$. The first factor is part of the norm $r^m_n(f)$ 
for some large $n$ and the second is one of the seminorms of $g$ in $\mathcal S$. The Proposition is proved and, as a bonus, we found out that 
$(f,g)\mapsto f\circ^B g$ extends to a bilinear {\it jointly continuous} mapping $:R^m_0\times\mathcal S\rightarrow\mathcal S$. 
\end{proof}

The class $C^\infty_{\text{\rm pol,u}}(\Xi)$ is indeed convenient. It has a very explicit definition and it contains all the polynomials in $x$ and $p$. 
It also contains the classical symbol spaces $S^m(\Xi):=\{f\in C^{\infty}(\Xi)\mid\vert(\partial^af)(\xi)\vert\le C_a<\xi>^{m-\vert a\vert},\;\forall 
a\in\mathbb N^{2N}\}$ for all $m$.

The magnetic Moyal algebra is large indeed, but many distributions, even with a good behaviour at infinity, are not inside. The one-rank 
projection $\vert u><u\vert$ is in $\mathcal L(\mathcal S)$ if and only if $u\in\mathcal S$. Thus, by Proposition \ref{imopMoy}, there are plenty 
of elements in $L^2(\Xi)$ not belonging to $\mathcal M(\Xi)$.

{\bf Acknowledgements:} Our understanding of the pseudodifferential theory without magnetic fields was influenced by unpublished lecture notes 
of Vladimir Georgescu and a  book in preparation of Gruia Arsu. We are grateful to George Nenciu and Mihai Pascu for useful discussions. Part of 
this work was done while the authors were visiting the Department of Theoretical Physics of the University of Geneva. We are greatful to 
Werner Amrein for his kind hospitality. We also acknowledge partial support from the EURROMMAT Pogramme (contract no. 
ICA1-CT-2000-70022), the CERES Programme (contract no. 38/2002) and the CNCSIS grant no. 33536/2003.

\textbf{Note added in proof:} After inserting the preprint of this paper on the mp-arc electronic archieve, T.A. Osborn brought to our attention the paper \textit{"Symplectic area, quantization, and dynamics in electromagnetic fields"}, J.  Math. Phys., \textbf{43} (2002), 756-788, by M.V. Karasev and T.A. Osborn and the more recent preprint \textit{"Quantum Magnetic Algebras and Magnetic Curvature"},  arXiv:quant-ph/0311053 by the same authors, where a gauge invariant quantization in the presence of an inhomogeneous electromagnetic tensor is developped in a way similar to ours. The above papers are concerned mainly with the geometric aspects related to this calculus and some interesting connections with grupoids. Our approach motivated by $C^*$-algebraic methods in spectral analysis for quantum Hamiltonians (see our previous paper: M. M\u antoiu and R. Purice, {\it The Algebra of Observables in a Magnetic Field}, Mathematical Results in Quantum 
Mechanics (Taxco, 2001), Contemporary Mathematics {\bf 307} (2002), Amer. Math. Soc., Providence, RI, 239-245) aims mainly to the analytic aspects of this quantization procedure (the main results of our Section 4).

\enddocument